\documentclass[superscriptaddress, prl, reprint, showpacs]{revtex4-1}

\usepackage[english]{babel}

\usepackage{amsmath}
\usepackage{amsthm}
\usepackage{amssymb}

\usepackage{color}
\usepackage{hyperref}
\usepackage{graphicx}
\usepackage{pdfpages}

\begin{document}


\title{Optical probe of ferroelectric order in bulk and thin film perovskite titanates}

\author{M.~R\"ossle}
\altaffiliation[present address:~]{University of Potsdam, Institute of Physics and Astronomy, Karl-Liebknecht-Strasse 24-25, D-14476 Potsdam, Germany}
\affiliation{University of Fribourg, Department of Physics and Fribourg Center for Nanomaterials, Chemin du Mus\'ee 3, CH-1700 Fribourg, Switzerland} 
\author{C.~N.~Wang} 
\affiliation{University of Fribourg, Department of Physics and Fribourg Center for Nanomaterials, Chemin du Mus\'ee 3, CH-1700 Fribourg, Switzerland} 
\author{P.~Marsik}
\affiliation{University of Fribourg, Department of Physics and Fribourg Center for Nanomaterials, Chemin du Mus\'ee 3, CH-1700 Fribourg, Switzerland} 
\author{M.~Yazdi-Rizi} 
\affiliation{University of Fribourg, Department of Physics and Fribourg Center for Nanomaterials, Chemin du Mus\'ee 3, CH-1700 Fribourg, Switzerland} 
\author{K.~W.~Kim}
\affiliation{University of Fribourg, Department of Physics and Fribourg Center for Nanomaterials, Chemin du Mus\'ee 3, CH-1700 Fribourg, Switzerland}
\affiliation{Department of Physics, Chungbuk National University, Cheongju 361-763, Korea} 
\author{A.~Dubroka} 
\affiliation{University of Fribourg, Department of Physics and Fribourg Center for Nanomaterials, Chemin du Mus\'ee 3, CH-1700 Fribourg, Switzerland} 
\affiliation{Department of Condensed Matter Physics, Faculty of Science, Masaryk University and Central European Institute of Technology, Kotl\'a\v rsk\'a 2, CZ-61137 Brno, Czech Republic}
\author{I.~Marozau} 
\affiliation{University of Fribourg, Department of Physics and Fribourg Center for Nanomaterials, Chemin du Mus\'ee 3, CH-1700 Fribourg, Switzerland} 
\author{C.~W.~Schneider} 
\affiliation{Paul Scherrer Institut, CH-5232 Villigen, Switzerland}
\author{J.~Huml\'i\v{c}ek}
\affiliation{Department of Condensed Matter Physics, Faculty of Science, Masaryk University and Central European Institute of Technology, Kotl\'a\v rsk\'a 2, CZ-61137 Brno, Czech Republic}
\author{D.~Baeriswyl}
\affiliation{University of Fribourg, Department of Physics and Fribourg Center for Nanomaterials, Chemin du Mus\'ee 3, CH-1700 Fribourg, Switzerland} 
\author{C.~Bernhard}
\email{christian.bernhard@unifr.ch}
\affiliation{University of Fribourg, Department of Physics and Fribourg Center for Nanomaterials, Chemin du Mus\'ee 3, CH-1700 Fribourg, Switzerland} 
 
\begin{abstract}
We have measured the temperature dependence of the direct band gap, $E_g$, in SrTi$^{16}$O$_3$ and BaTiO$_3$ and related materials with quantum-paraelectric and ferroelectric properties using optical spectroscopy. We show that $E_g$ exhibits an anomalous temperature dependence with pronounced changes in the vicinity of the ferroelectric transition that can be accounted for in terms of the Fr\"ohlich electron-phonon interaction with an optical phonon mode, the so-called soft mode. We demonstrate that these characteristic changes of $E_g$ can be readily detected even in very thin films of SrTi$^{16}$O$_3$ with a strain-induced ferroelectric order. Optical spectroscopy thus can be used as a relatively simple but sensitive probe of ferroelectric order in very thin films of these titanates and probably also in subsequent multilayers and devices. 
\end{abstract}

\pacs{78.20.-e, 77.84.-s, 77.55.-g, 63.20.kd}

\maketitle



Ferroelectric (FE) materials have important applications in sensors and electronic devices, for example in dynamic random access memories \cite{Ferroelectrics2007, Setter2006, Bibes2011}. The growth and characterization of very thin FE layers is therefore an active field of research. In FE thin films the polarization and the transition temperature, $T_{\mathrm{Curie}}$, can be severely reduced below their bulk values or even entirely suppressed due to so-called ``dead layers'' that may arise from poorly screened depolarization fields at surfaces and interfaces or from strain and defects \cite{Ferroelectrics2007, Setter2006, Bibes2011}. On the other hand, in SrTi$^{16}$O$_3$, which in its bulk form is a so-called quantum paraelectric material for which the FE order is suppressed by quantum lattice fluctuations, a FE order can be induced in thin films that are strained by lattice matching to the substrate \cite{Pertsev2000, Haeni2004}. For such thin films and especially for subsequent complex heterostructures with FE layers, it is therefore very important to have a readily accessible technique that enables one to identifiy the FE order. Direct capacitance or impedance spectroscopy measurements of the FE order on such thin films are often complicated and hindered by parasitic charges forming at interfaces and electrodes that can obscure the FE contribution \cite{Lunkenheimer2002, Langenberg2012}. Dedicated techniques like piezo-response force microscopy \cite{Kalinin2006}, synchrotron x-ray diffraction \cite{Streiffer2002, Fong2004} or ultraviolet Raman spectroscopy \cite{Tenne2006} have therefore been used to detect the FE order in very thin films. In the following we show that standard optical spectroscopy measurements of the temperature ($T$) dependence of the direct band gap, $E_g$, can be used as an alternative and very effective tool to identify such a FE transition. This is demonstrated for the case of bulk SrTi$^{16}$O$_3$ and BaTiO$_3$ crystals as well as for strained SrTi$^{16}$O$_3$ thin films.

 
SrTi$^{16}$O$_3$ (STO) and KTaO$_3$ crystals were purchased from Crystec \cite{Crystec} and a BaTiO$_3$ crystal from SurfaceNet \cite{SurfaceNet}, respectively. A polycrystalline CaTiO$_3$ sample was made via solid-state sintering. A $^{18}$O exchanged SrTi$^{18}$O$_3$ crystal was prepared as described in Ref.~\cite{Wang2001}. STO thin films on DyScO$_3$ and (LaAlO$_3$)$_{0.3}$--(SrAlTaO$_6$)$_{0.7}$ (LSAT) substrates were prepared by pulsed laser deposition with in situ reflection high energy electron diffraction growth control. The substrate temperature was 900$^{\circ}$C, the oxygen pressure $p(\mathrm{O}_2) = 0.11$~mbar and we used a 248~nm excimer laser with a fluence of 1.5~$\frac{\mathrm{J}}{\mathrm{cm}^2}$. Subsequent to the growth, the films were slowly cooled in 1~bar oxygen. 

Spectroscopic ellipsometry measurements were performed in the near-infrared to ultra-violet range of 0.5 -- 6.5~eV with a commercial ellipsometer (Woollam VASE \cite{Woollam}) equipped with a UHV cryostat for a temperature range of $4\;\mathrm{K} < T \leq 700$~K. For the far-infrared (FIR) and mid-infrared measurements we used a home-built setup as described in Ref. \cite{Bernhard2004}. FIR reflectivity measurements were performed with the home-built setup described in Ref.~\cite{Kim2010}. The modeling of the ellipsometry data (surface roughness correction for crystals and substrate correction for thin films) was done with the Woollam VASE software ~\cite{Woollam}.


\begin{figure}[!ht]
  \centering
    \includegraphics[width = \columnwidth]{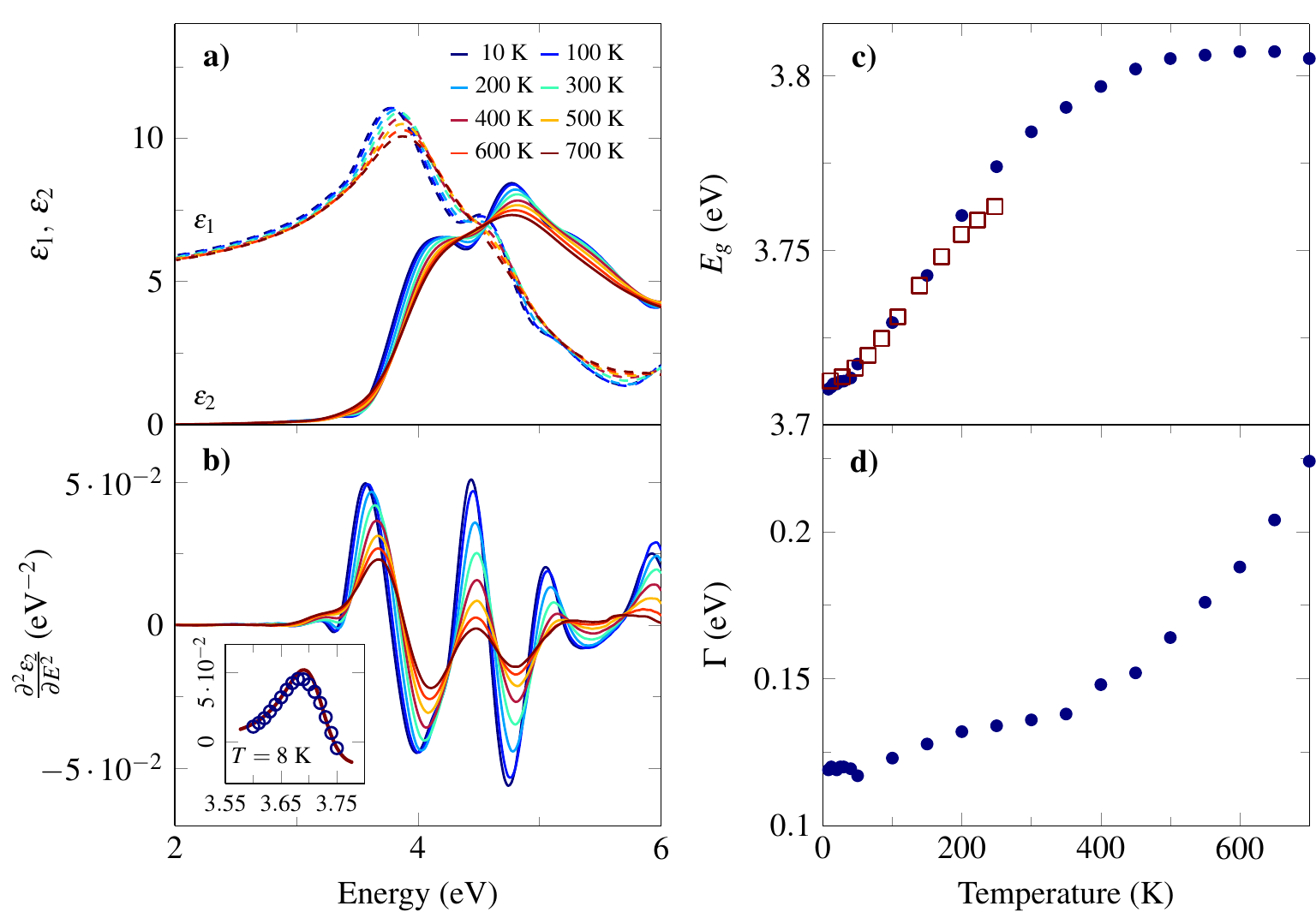}
  \caption{(color online) \textbf{a)} $T$-dependent spectra of the dielectric function, $\varepsilon = \varepsilon_1 + \mathrm{i} \varepsilon_2$, of STO. \textbf{b)} Corresponding second derivative spectra of $\varepsilon_2$. $T$-dependence of \textbf{c)} the energy, $E_g$, and \textbf{d)} the broadening, $\Gamma$ of the direct gap (full circles) as obtained by fitting with the second derivative of Eq.~\eqref{eq:1}. Open squares in \textbf{c)} show the contribution of the Fr\"ohlich electron-phonon interaction to the $T$-dependence of $E_g$ below 250~K as estimated with Eq.~\eqref{eq:2}. The inset in \textbf{b)} shows a representative fit of the lowest direct interband transition (solid line) using Eq.~\eqref{eq:1} to the second derivative of the experimental data (open symbols) at $T=8$~K.}
  \label{fig:1}
\end{figure}

Figure~\ref{fig:1}(a) shows the $T$-dependent spectra from 2 -- 6~eV of the real and imaginary parts of the dielectric function, $\varepsilon = \varepsilon_1 + \mathrm{i} \varepsilon_2$, of bulk STO. They contain four prominent bands centered at 3.7, 4.2, 5.1, and about 6~eV which correspond to direct interband transitions between the O-$2p$ states in the valence band and the Ti-$3d$ states in the conduction band \cite{Cardona1965, Benthem2001}. The indirect band gap at 3.25~eV yields a very weak absorption band that is barely visible in the spectra and therefore does not concern the following discussion of the $T$-dependence of the direct band gap, $E_g$, at about 3.7~eV. Figure~\ref{fig:1}(b) shows the spectra of the second derivative of $\varepsilon_2$ as obtained with a Savitzky-Golay smoothing procedure \cite{Savitzky1964}. They have been fitted with the second derivative of the function 
\begin{equation}
  \varepsilon(\omega) = C - S \cdot e^{\mathrm{i} 3\pi/2} ( \omega - E + \mathrm{i}\Gamma )^{1/2}\;\;,\label{eq:1}
\end{equation}
which describes the response in the vicinity of a direct absorption edge due to the singularity in the joint density of states at a three-dimensional so-called $M_0$ critical point \cite{Yu2005, Lautenschlager1987}. Here, the parameter $C$ represents a constant background, $S$ is the amplitude, $\Gamma$ the broadening, and $E$ corresponds to the energy threshold of the associated transition which in this case is the energy of the band gap, $E_g$. The obtained $T$-dependence of $E_g$ and $\Gamma$ are displayed in Figs.~\ref{fig:1}(c) and (d), respectively. The evolution of $E_g$ is rather anomalous, it exhibits a sizeable increases from $\sim3.71$~eV at 10~K to 3.77~eV at 300~K before it reaches a maximum at $\sim3.81$~eV around 500~K and decreases again toward higher $T$. We note that this characteristic $T$-dependent shift of the interband transition is visible in the bare spectra of $\varepsilon$, i.e. it can be readily identified even without the quantitative analysis using Eq.~\eqref{eq:1}. A similar trend below 300~K was also previously reported in Ref.~\cite{Trepakov2009}. 

The magnitude of $E_g$ in semiconductors and insulators is well known to undergo sizeable changes with $T$, usually to lower energy, like in Si, Ge or GaAs \cite{Lautenschlager1985, Olguin2002}, but in some cases as in PbS also to higher energy \cite{Keffer1968}. Part of this $T$-dependence is accounted for by the thermal expansion of the lattice, but a sizeable contribution can also arise from the electron-phonon interaction \cite{Allen1981, Allen1983}. The latter is caused by the deformation of the electronic potentials due to the dynamic atomic displacements. Its sign is determined by the lattice structure and the electronic states forming the valence and conduction bands. Its magnitude depends on the amplitude of the atomic displacement, $u$, which in the harmonic approximation is related to the effective mass of the atoms, $\mu$, the eigenfrequency of the phonon mode, $\omega$, and the Bose-Einstein factor, $n_B = \left(e^{\hbar \omega/k_B T} - 1 \right)^{-1}$, according to $\left<u^2\right> = \frac{\hbar}{2 \mu \omega}( 1 + 2 n_B )$ where $\left< \right>$ means a thermal average. The resulting shift of the band gap, $\Delta E_g$, is nearly constant at $k_B T \ll \hbar\omega$ where it is dominated by the quantum lattice fluctuations, it starts to increase around $k_BT\approx\hbar\omega$ and is proportional to $T$ at $k_B T \gg \hbar\omega$ \cite{Cardona2005a}. As observed in some chalcopyrites, the $T$-dependence of $E_g$ may also exhibit a sign change if the contributions due to phonons with low and high eigenfrequencies have opposite signs \cite{Bhosale2012}. 

In polar or partially ionic crystals, an additional contribution to the electron-phonon interaction arises from the long wavelength longitudinal optical (LO) phonons which give rise to a macroscopic polarization that can be described in terms of the Fr\"ohlich interaction \cite{Yu2005}. Its contribution to the shift of the band gap is \cite{Allen1981, Fan1951} 
\begin{equation}
\Delta E_g = -A \cdot \left(\varepsilon_{\infty}^{-1}-\varepsilon_0^{-1}\right)\left( 1 + 2 n_B \right) \;\;,\label{eq:2}
\end{equation}
where $\varepsilon_{\infty}$ and $\varepsilon_0$ are the dielectric constants at energies well above and below the phonon range, respectively, and $A$ is a $T$-independent prefactor that depends on material parameters such as effective mass and lattice constant. It is important to notice that at $k_B T \ll \hbar\omega$, where $n_B$ is nearly constant, the $T$-dependence of $\Delta E_g$ is determined by that of $\varepsilon_0$.

In the following we provide evidence that in STO and related titanates this Fr\"ohlich interaction governs the $T$-dependence of $E_g$ below about 250~K. In STO the thermal expansion of the lattice has a regular $T$-dependence \cite{Liu1997, Loetzsch2010} and thus should contribute to a decrease of $E_g$ as the lattice expands with increasing $T$. This assumption is confirmed by band structure calculations which predict that $E_g$ increases with pressure (or decreasing lattice constant) \cite{Ghebouli2009}. The structural phase transitions at 105~K from a cubic to a tetragonal state and around 65~K toward an orthorhombic phase also do not have any noticeable effect on $E_g$. This suggests that the unusual $T$-dependence of $E_g$ in STO is caused by the electron-phonon interaction. STO is indeed well known for its anomalous lattice dynamical properties. It is a so-called quantum paraelectric material with an incipient FE state below about 35~K \cite{Lytle1964} that is suppressed by the quantum lattice fluctuations \cite{Muller1979}. This anomalous behavior is closely related to the softening of a transverse optical (TO) phonon which involves a polar displacement of the Ti ion away from the central position of the octahedron formed by the neighboring oxygen ions. The eigenfrequency of this so-called ``soft mode'', $\omega_{\mathrm{TO}}$, decreases from about 95~cm$^{-1}$ at 300~K to about 15~cm$^{-1}$ at low $T$ \cite{Vogt1995}. According to the Lyddane-Sachs-Teller relation, $\frac{\varepsilon_0}{\varepsilon_{\infty}} = C\cdot\frac{\omega_{\mathrm{LO}}^2}{\omega_{\mathrm{TO}}^2}$ this soft TO mode gives rise to a strong increase of $\varepsilon_0 = 390$ at 300~K to $\varepsilon_0 = 14.000$ at low $T$ \cite{Servoin1980, Sirenko2000}. The open squares in Fig.~\ref{fig:1}(c) show that this divergence of $\varepsilon_0$ toward low $T$ and the subsequent increase of the Fr\"ohlich interaction according to Eq.~\eqref{eq:2} account well for the renormalisation of $E_g$ below about 250~K. The fit to the $E_g$ data yields a vertical offset of $E_g(T=0~\mathrm{K}) = 4.15\pm0.02$~eV and a prefactor of $A = 2.26\pm0.12$~eV. The prefactor $C=2.52\pm0.12$ in the Lyddane-Sachs-Teller relation, which accounts for the $T$-independent contribution of the two other IR-active phonon modes at 175 and 540~cm$^{-1}$, and $\varepsilon_{\infty} = 5.1$ and the eigenfrequency of the LO mode, $\omega_{\mathrm{LO}} = 788$~cm$^{-1}$, have been obtained from the ellipsometry data. The latter have only a very weak $T$-dependence as shown in Ref.~\cite{SOM} that does not have a noticeable effect on our estimate of the Fr\"ohlich interaction.

\begin{figure}[!ht]
  \centering
    \includegraphics[width = \columnwidth]{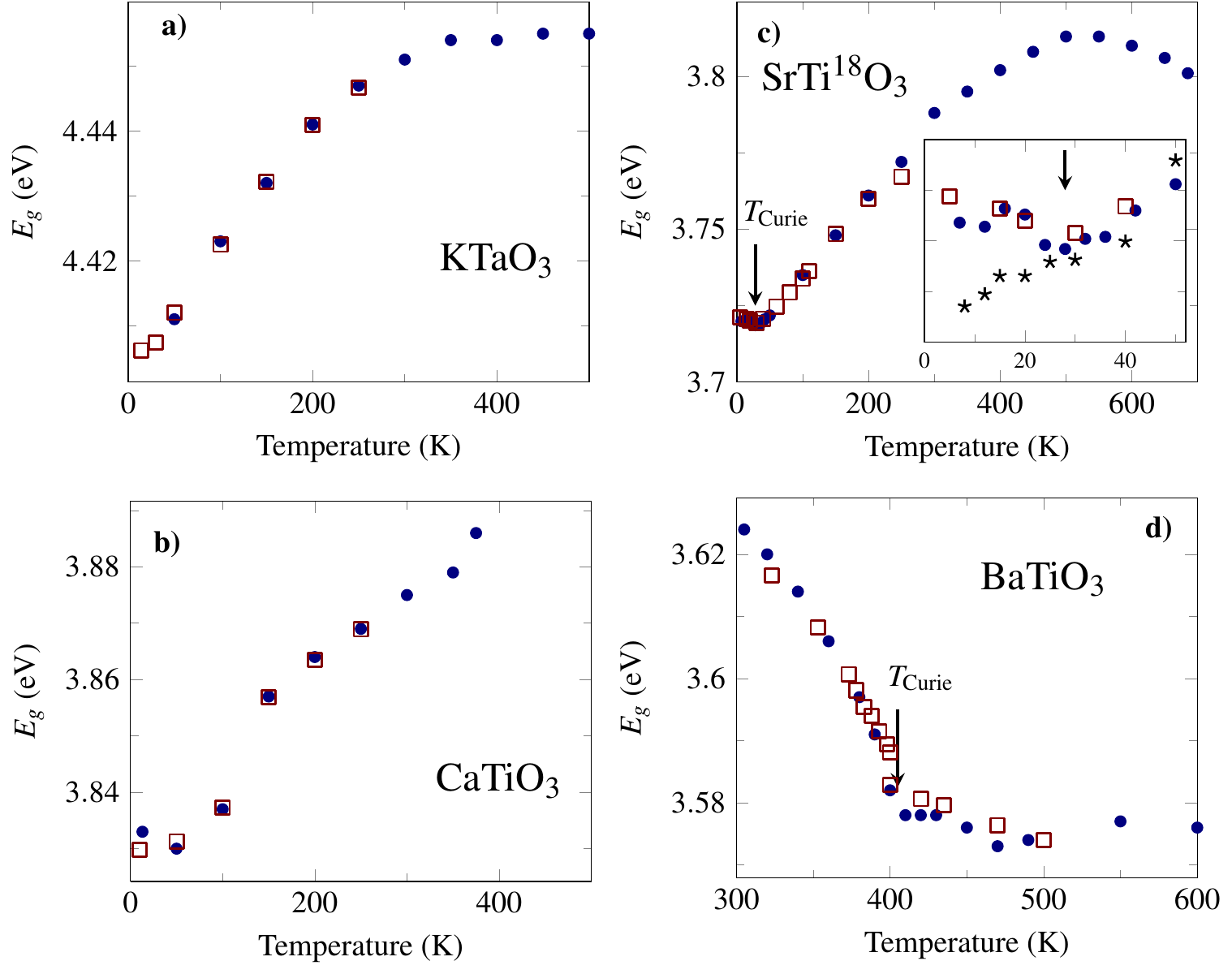}
  \caption{(color online) $T$-dependence of $E_g$ (full circles) in \textbf{a)} KTaO$_3$ and \textbf{b)} CaTiO$_3$. Open squares show the contribution of the Fr\"ohlich-interaction as obtained with equation~\eqref{eq:2}. \textbf{c)} Corresponding data (full circles) and calculations (open squares) for SrTi$^{18}$O$_3$ with a FE transition at $T_{\mathrm{Curie}} = 26$~K as marked by the arrow. Inset: Magnification of the low-$T$ regime with the rescaled data of STO (stars) shown for comparison. \textbf{d)} Corresponding data and calculations for FE BaTiO$_3$ with $T_{\mathrm{Curie}} = 405$~K. }
  \label{fig:2}
\end{figure}

Figure~\ref{fig:2}(a) shows the $T$-dependence of $E_g$ in KTaO$_3$ which is also a quantum paraelectric material with a soft TO phonon mode \cite{Perry1967}. Similar to STO, the value of $E_g$, as derived from the optical spectra, exhibits a strong increase with $T$ up to about 300~K. The open squares show that the Fr\"ohlich-interaction accounts once more well for the $T$-dependent renormalisation of $E_g$ below 250~K. As input parameters we used $\varepsilon_{\infty} = 4.7$ and the $T$-dependence of $\omega_{\mathrm{TO}}$ that was measured with FIR ellipsometry as outlined in the Supplementary Material \cite{SOM}. The value for $\omega_{\mathrm{LO}} = 833$~cm$^{-1}$ has been taken from Ref.~\cite{Fleury1968}. The fit using equation~\eqref{eq:2} yields a vertical offset of $E_g(T=0~\mathrm{K}) = 4.87\pm0.01$~eV, $A = 2.24\pm0.06$~eV, and $C = 2.92\pm0.07$. The corresponding data and fits for CaTiO$_3$ with $C=1.89\pm0.08$, $E_g(T=0~\mathrm{K}) = 4.35\pm0.03$~eV, and $A = 3.30\pm0.15$~eV are displayed in Fig.~\ref{fig:2}(b). They confirm that the soft mode behavior and the resulting divergence of $\varepsilon_0$ towards low $T$ are governing the renormalisation of $E_g$. To the best of our knowledge, this is the first time that the important role of the Fr\"ohlich electron-phonon interaction in the anomalous $T$-dependence of $E_g$ in these quantum paraelectric materials has been demonstrated.  

In return, the characteristic $T$-dependence of $E_g$ can be used to monitor the soft mode behavior and to identify a possible FE transition in these materials. We demonstrate this first for the case of oxygen isotope substituted SrTi$^{18}$O$_3$ for which a weak FE order that is stabilized by the reduction of the lattice quantum fluctuations develops below $T_{\mathrm{Curie}}\approx 26$~K \cite{Itoh1999}. Figure \ref{fig:2}(c) shows that the $T$-dependence of $E_g$ exhibits indeed a weak, yet clearly noticeable anomaly at $T_{\mathrm{Curie}}\approx 26$~K. This anomaly is marked by an arrow and is more clearly shown in the inset which compares the low-$T$ behavior of $E_g$ in  SrTi$^{18}$O$_3$ (full circles) with the one of the rescaled $E_g$ (by a factor of 1.0015) of SrTi$^{16}$O$_3$ (stars). Shown by the open squares is the contribution of the Fr\"ohlich interaction as obtained using Eq.~\eqref{eq:2} with $\omega_{\mathrm{LO}} = 765$~cm$^{-1}$, $\varepsilon_{\infty} = 5.1$, $C=1.88\pm0.05$, and the $T$-dependence of $\omega_{\mathrm{TO}}$ as measured with FIR spectroscopy. The latter exhibits an anomalous hardening below $T_{\mathrm{Curie}}$ as is shown in the Supplemental Material \cite{SOM}. The fit yields $E_g(T=0~\mathrm{K}) = 4.34\pm0.02$~eV and $A = 3.16\pm0.10$~eV and reproduces the $T$-dependence of $E_g$ and also the anomaly around $T_{\mathrm{Curie}}\approx 26$~K rather well. 

It has been predicted that the static displacement, $u$, that occurs below the FE transition will also contribute to the hardening of $E_g$ \cite{Berger2011}. It reduces the symmetry of the lattice and thereby can introduce a mixing of the oxygen and titanium related bands that leads to a repulsion of the valence and conduction bands which increases $E_g$. Nevertheless, the good agreement in Fig~\ref{fig:2}(c) between the measured $E_g$ values and the estimated contribution of the Fr\"ohlich interaction suggests that the latter accounts for a sizeable and possibly even the major part of the hardening of $E_g$ in the FE state.

In Figure~\ref{fig:2}(d) we show that similar arguments likely even apply for the case of BaTiO$_3$ for which the formation of a FE state with a large polarization below $T_{\mathrm{Curie}}=405$~K gives rise to a much stronger anomaly in the $T$-dependence of $E_g$ than in SrTi$^{18}$O$_3$. Whereas $E_g$ is only weakly $T$-dependent in the paraelectric state at $T>T_{\mathrm{Curie}}$, right below $T_{\mathrm{Curie}}$ it exhibits a steep increase that is about 50 times stronger than the one in SrTi$^{18}$O$_3$. In BaTiO$_3$ the situation is somewhat complicated by the fact that there exist two soft TO modes that both contribute to the divergence of $\varepsilon_0$ at $T_{\mathrm{Curie}}$ \cite{Hlinka2008, Ponomareva2008}. The additional low frequency mode (the so-called central mode) accounts here for the order-disorder component of the FE transition which apparently is of a mixed character. Accordingly, we explicitly included the $T$-dependence of both soft TO modes in the Lyddane-Sachs-Teller relation using $\omega_{\mathrm{TO}}$ as reported in Refs.~\cite{Vogt1982, Ponomareva2008, Hlinka2008}. The values of $C=4.78\pm0.20$, $\omega_{\mathrm{LO}}(300~\mathrm{K}) = 716$~cm$^{-1}$ and $\varepsilon_{\infty}=5.3$ were once more obtained from our IR ellipsometry data. The open squares in Figure~\ref{fig:2}(d) show that the large anomaly of $E_g$ in BaTiO$_3$ can be reasonably well reproduced in terms of the Fr\"ohlich interaction with the fit parameters $E_g(T=0~\mathrm{K})=3.72\pm0.01$~eV and $A = 0.63\pm0.01$~eV that are not too far from the ones in STO. We notice that the agreement could be further improved by including additional contributions from the coupling to the other phonon modes (which do not soften toward $T_{\mathrm{Curie}}$) and the thermal lattice expansion. These can become rather sizeable in the vicinity of $T_{\mathrm{Curie}}=405$~K and may for example account for the rather weak $T$-dependence of $E_g$ in the paraelectric state.     

\begin{figure}[!ht]
  \centering
    \includegraphics[width = \columnwidth]{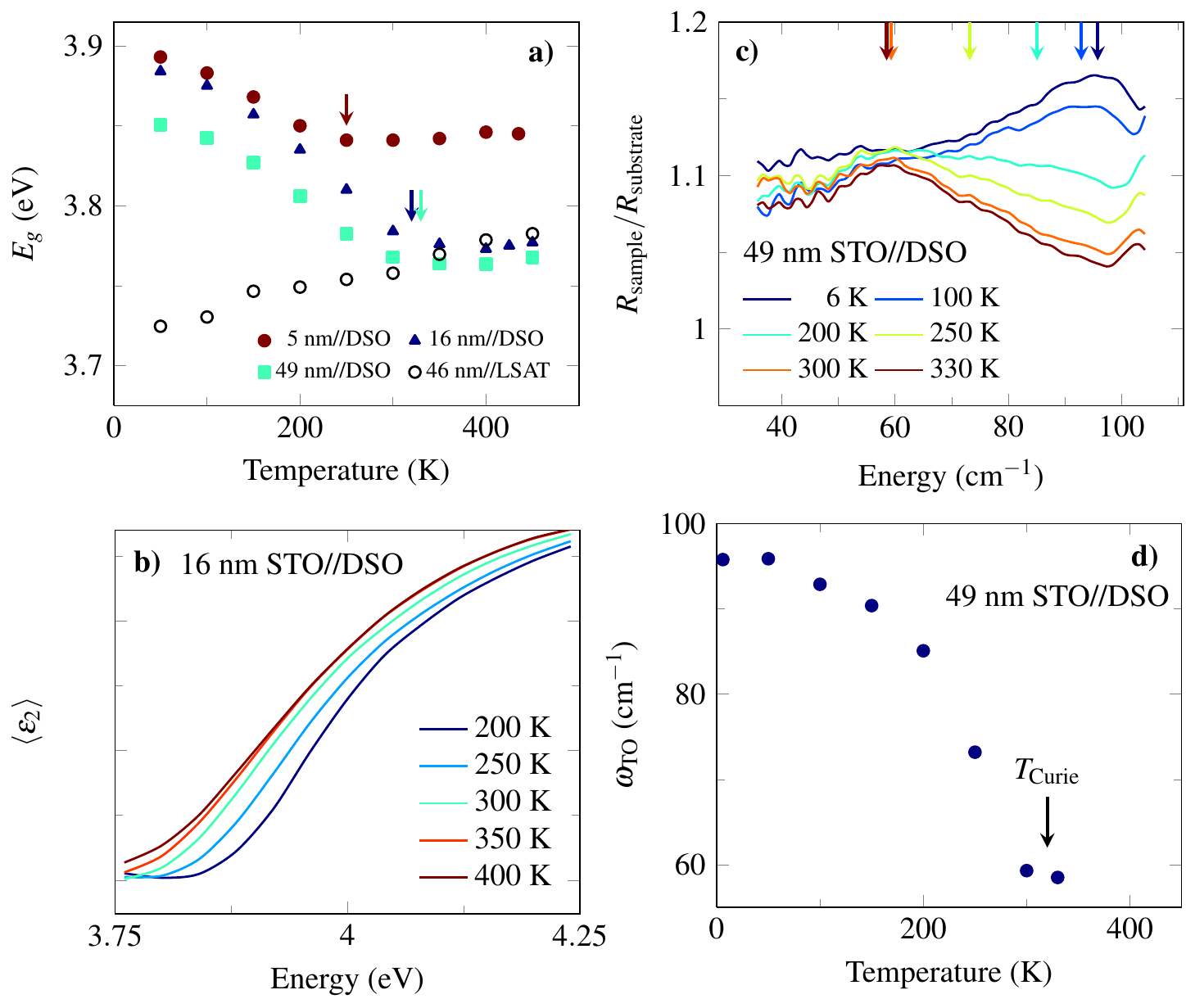}
  \caption{(color online) \textbf{a)} $T$-dependence of $E_g$ in strained STO thin films on DyScO$_3$ and LSAT substrates. Arrows mark the anomalies at $T_{\mathrm{Curie}}$ for the films on DyScO$_3$. \textbf{b)} Evolution of the as-measured pseudo-dielectric function, $\left<\varepsilon_2\right>$, of the 16~nm STO film on DyScO$_3$. \textbf{c)} Ratio of the FIR reflectivity of sample and bare substrate for the 49~nm thick STO film on DyScO$_3$. The arrows indicate the soft mode frequency $\omega_{\mathrm{TO}}$ for which the $T$-dependence is detailed in \textbf{d)}.}
  \label{fig:3}
\end{figure}

Last but not least, we show in  Fig.~\ref{fig:3} that the anomaly in the $T$-dependence of $E_g$ can be used to identify the FE order in STO thin films. It is well known that a FE order can be induced in STO thin films that are grown on substrates which introduce a strong tensile strain \cite{Haeni2004}. For orthorhombic, (110) oriented DyScO$_3$ substrates the average mismatch with respect to the lattice parameters of bulk SrTiO$_3$ is about +1\%. For fully lattice matched STO this amounts to a sizeable tensile stress of $\sim1$~GPa and an induced FE transition at $T_{\mathrm{Curie}}\sim300$~K. This FE state has been previously identified with piezo-force scanning microscopy \cite{Kalinin2006}, synchrotron x-ray diffraction \cite{Streiffer2002}, and more recently with THz and FIR spectroscopy studies of the $T$-dependence of the soft TO mode of STO \cite{Nuzhnyy2009}. 

Figure~\ref{fig:3}(a) shows that the strain-induced FE order in these STO thin films can be also conveniently identified with optical spectroscopy. The solid symbols display the $T$-dependence of $E_g$ for a series of STO films on DyScO$_3$ substrates with thicknesses of 49, 16, and 5~nm, respectively. The characteristic anomaly, as indicated by the arrows, serves as a clear fingerprint of the FE transition at $T_{\mathrm{Curie}}\sim330$, 320 and 250~K, respectively. For the case of the 49~nm thick STO film we have confirmed the FE transition at $T_{\mathrm{Curie}}=330$~K with far-infrared reflectivity measurements which directly probe the $T$-dependence of the soft TO mode as shown in Figs.~\ref{fig:3}(c)-(d). For the thinner films these far-infrared data are increasingly difficult to interpret since the signal from the  STO phonons becomes much weaker than the one from the DyScO$_3$ substrate (which also exhibits a considerable $T$-dependence). On the other hand, Fig.~\ref{fig:3}(b) shows for the 16~nm thick STO film that the FE transition can be readily identified in the as measured spectra of the pseudo-dielectric function, $\left<\varepsilon_2\right>$ due to the blue shift of the corresponding interband transition below $T_{\mathrm{Curie}}\sim320$~K. The enhanced sensitivity of the optical spectra to the properties of such thin films is owed to the penetration depth of the light which is much shorter in the VIS-UV range ($\sim100$~nm) than in the FIR range ($>1000$~nm). Finally, Fig.~\ref{fig:3}(a) also shows our optical data of a STO thin film grown on a LSAT substrate which exerts a weak compressive strain that does not induce a FE transition \cite{Haeni2004}. Here the $T$-dependence of $E_g$ exhibits indeed no sign of an anomaly and is similar to the one in bulk STO. 

In summary, we have shown that the anomalous $T$-dependent shift of the direct band gap of STO is strongly affected and likely even dominated by the Fr\"ohlich electron-phonon interaction with the so-called soft mode that is at the heart of its quantum paraelectric properties. We have also demonstrated that the optical band gap measurements can therefore be used as a very sensitive and efficient tool to search for a ferroelectric order in related bulk and thin film materials. In particular, we have shown that the strain-induced ferroelectric order in STO films on a DyScO$_3$ substrate can be readily detected even in films that are only 5~nm thick. This well established optical technique, which is accessible in many laboratories, could be rather useful to search for ferroelectric transitions in other bulk and thin film materials, in particular, in complex heterostructures and devices with buried ferroelectric layers. It will also be interesting to study relaxor ferroelectric, improper ferroelectric or even multiferroic materials to investigate whether they exhibit a similar relationship between the enhancement of the dielectric constant and the renormalisation of the direct band gap due to the Fr\"ohlich-type electron-phonon interaction.  


\begin{acknowledgments}
This work is supported by the Schweizer Nationalfonds (SNF) grant 200020-140225 and the project ``CEITEC-Central European Institute of Technology'' (CZ.1.05/1.1.00/02.0068). We thank K. Conder for his help in preparing the isotope exchanged SrTi$^{18}$O$_3$ crystal and B. Doggett for his contribution to the PLD thin film growth. We acknowledge stimulating discussions with Manuel Cardona and Eugene Kotomin.
\end{acknowledgments}


\begin{thebibliography}{48}%
\makeatletter
\providecommand \@ifxundefined [1]{%
 \@ifx{#1\undefined}
}%
\providecommand \@ifnum [1]{%
 \ifnum #1\expandafter \@firstoftwo
 \else \expandafter \@secondoftwo
 \fi
}%
\providecommand \@ifx [1]{%
 \ifx #1\expandafter \@firstoftwo
 \else \expandafter \@secondoftwo
 \fi
}%
\providecommand \natexlab [1]{#1}%
\providecommand \enquote  [1]{``#1''}%
\providecommand \bibnamefont  [1]{#1}%
\providecommand \bibfnamefont [1]{#1}%
\providecommand \citenamefont [1]{#1}%
\providecommand \href@noop [0]{\@secondoftwo}%
\providecommand \href [0]{\begingroup \@sanitize@url \@href}%
\providecommand \@href[1]{\@@startlink{#1}\@@href}%
\providecommand \@@href[1]{\endgroup#1\@@endlink}%
\providecommand \@sanitize@url [0]{\catcode `\\12\catcode `\$12\catcode
  `\&12\catcode `\#12\catcode `\^12\catcode `\_12\catcode `\%12\relax}%
\providecommand \@@startlink[1]{}%
\providecommand \@@endlink[0]{}%
\providecommand \url  [0]{\begingroup\@sanitize@url \@url }%
\providecommand \@url [1]{\endgroup\@href {#1}{\urlprefix }}%
\providecommand \urlprefix  [0]{URL }%
\providecommand \Eprint [0]{\href }%
\providecommand \doibase [0]{http://dx.doi.org/}%
\providecommand \selectlanguage [0]{\@gobble}%
\providecommand \bibinfo  [0]{\@secondoftwo}%
\providecommand \bibfield  [0]{\@secondoftwo}%
\providecommand \translation [1]{[#1]}%
\providecommand \BibitemOpen [0]{}%
\providecommand \bibitemStop [0]{}%
\providecommand \bibitemNoStop [0]{.\EOS\space}%
\providecommand \EOS [0]{\spacefactor3000\relax}%
\providecommand \BibitemShut  [1]{\csname bibitem#1\endcsname}%
\let\auto@bib@innerbib\@empty
\bibitem [{\citenamefont {Rabe}\ \emph {et~al.}(2007)\citenamefont {Rabe},
  \citenamefont {Ahn},\ and\ \citenamefont {Triscone}}]{Ferroelectrics2007}%
  \BibitemOpen
  \bibinfo {editor} {\bibfnamefont {K.~M.}\ \bibnamefont {Rabe}}, \bibinfo
  {editor} {\bibfnamefont {C.~H.}\ \bibnamefont {Ahn}}, \ and\ \bibinfo
  {editor} {\bibfnamefont {J.-M.}\ \bibnamefont {Triscone}},\ eds.,\ \href@noop
  {} {\emph {\bibinfo {title} {{P}hysics of {F}erroelectrics -- {A} {M}odern
  {P}erspective}}}\ (\bibinfo  {publisher} {Springer-Verlag},\ \bibinfo {year}
  {2007})\BibitemShut {NoStop}%
\bibitem [{\citenamefont {{S}etter}\ \emph {et~al.}(2006)\citenamefont
  {{S}etter}, \citenamefont {{D}amjanovic}, \citenamefont {{E}ng},
  \citenamefont {{F}ox}, \citenamefont {{G}evorgian}, \citenamefont {{H}ong},
  \citenamefont {{K}ingon}, \citenamefont {{K}ohlstedt}, \citenamefont
  {{P}ark}, \citenamefont {{S}tephenson}, \citenamefont {{S}tolitchnov},
  \citenamefont {{T}agantsev}, \citenamefont {{T}aylor}, \citenamefont
  {{Y}amada},\ and\ \citenamefont {{S}treiffer}}]{Setter2006}%
  \BibitemOpen
  \bibfield  {author} {\bibinfo {author} {\bibfnamefont {N.}~\bibnamefont
  {{S}etter}}, \bibinfo {author} {\bibfnamefont {D.}~\bibnamefont
  {{D}amjanovic}}, \bibinfo {author} {\bibfnamefont {L.}~\bibnamefont {{E}ng}},
  \bibinfo {author} {\bibfnamefont {G.}~\bibnamefont {{F}ox}}, \bibinfo
  {author} {\bibfnamefont {S.}~\bibnamefont {{G}evorgian}}, \bibinfo {author}
  {\bibfnamefont {S.}~\bibnamefont {{H}ong}}, \bibinfo {author} {\bibfnamefont
  {A.}~\bibnamefont {{K}ingon}}, \bibinfo {author} {\bibfnamefont
  {H.}~\bibnamefont {{K}ohlstedt}}, \bibinfo {author} {\bibfnamefont {N.~Y.}\
  \bibnamefont {{P}ark}}, \bibinfo {author} {\bibfnamefont {G.~B.}\
  \bibnamefont {{S}tephenson}}, \bibinfo {author} {\bibfnamefont
  {I.}~\bibnamefont {{S}tolitchnov}}, \bibinfo {author} {\bibfnamefont {A.~K.}\
  \bibnamefont {{T}agantsev}}, \bibinfo {author} {\bibfnamefont {D.~V.}\
  \bibnamefont {{T}aylor}}, \bibinfo {author} {\bibfnamefont {T.}~\bibnamefont
  {{Y}amada}}, \ and\ \bibinfo {author} {\bibfnamefont {S.}~\bibnamefont
  {{S}treiffer}},\ }\href {\doibase 10.1063/1.2336999} {\bibfield  {journal}
  {\bibinfo  {journal} {{J}. {A}ppl. {P}hys.}\ }\textbf {\bibinfo {volume}
  {100}},\ \bibinfo {eid} {051606} (\bibinfo {year} {2006})}\BibitemShut
  {NoStop}%
\bibitem [{\citenamefont {{B}ibes}\ \emph {et~al.}(2011)\citenamefont
  {{B}ibes}, \citenamefont {{V}illegas},\ and\ \citenamefont
  {{B}arth\'el\'emy}}]{Bibes2011}%
  \BibitemOpen
  \bibfield  {author} {\bibinfo {author} {\bibfnamefont {M.}~\bibnamefont
  {{B}ibes}}, \bibinfo {author} {\bibfnamefont {J.~E.}\ \bibnamefont
  {{V}illegas}}, \ and\ \bibinfo {author} {\bibfnamefont {A.}~\bibnamefont
  {{B}arth\'el\'emy}},\ }\href
  {http://www.informaworld.com/10.1080/00018732.2010.534865} {\bibfield
  {journal} {\bibinfo  {journal} {{A}dv. {P}hys.}\ }\textbf {\bibinfo {volume}
  {60}},\ \bibinfo {pages} {5} (\bibinfo {year} {2011})}\BibitemShut {NoStop}%
\bibitem [{\citenamefont {{P}ertsev}\ \emph {et~al.}(2000)\citenamefont
  {{P}ertsev}, \citenamefont {{T}agantsev},\ and\ \citenamefont
  {{S}etter}}]{Pertsev2000}%
  \BibitemOpen
  \bibfield  {author} {\bibinfo {author} {\bibfnamefont {N.~A.}\ \bibnamefont
  {{P}ertsev}}, \bibinfo {author} {\bibfnamefont {A.~K.}\ \bibnamefont
  {{T}agantsev}}, \ and\ \bibinfo {author} {\bibfnamefont {N.}~\bibnamefont
  {{S}etter}},\ }\href {\doibase 10.1103/PhysRevB.61.R825} {\bibfield
  {journal} {\bibinfo  {journal} {{P}hys. {R}ev. {B}}\ }\textbf {\bibinfo
  {volume} {61}},\ \bibinfo {pages} {R825} (\bibinfo {year}
  {2000})}\BibitemShut {NoStop}%
\bibitem [{\citenamefont {{H}aeni}\ \emph {et~al.}(2004)\citenamefont
  {{H}aeni}, \citenamefont {{I}rvin}, \citenamefont {{C}hang}, \citenamefont
  {{U}ecker}, \citenamefont {{R}eiche}, \citenamefont {{L}i}, \citenamefont
  {{C}houdhury}, \citenamefont {{T}ian}, \citenamefont {{H}awley},
  \citenamefont {{C}raigo}, \citenamefont {{T}agantsev}, \citenamefont {{P}an},
  \citenamefont {{S}treiffer}, \citenamefont {{C}hen}, \citenamefont
  {{K}irchoefer}, \citenamefont {{L}evy},\ and\ \citenamefont
  {{S}chlom}}]{Haeni2004}%
  \BibitemOpen
  \bibfield  {author} {\bibinfo {author} {\bibfnamefont {J.~H.}\ \bibnamefont
  {{H}aeni}}, \bibinfo {author} {\bibfnamefont {P.}~\bibnamefont {{I}rvin}},
  \bibinfo {author} {\bibfnamefont {W.}~\bibnamefont {{C}hang}}, \bibinfo
  {author} {\bibfnamefont {R.}~\bibnamefont {{U}ecker}}, \bibinfo {author}
  {\bibfnamefont {P.}~\bibnamefont {{R}eiche}}, \bibinfo {author}
  {\bibfnamefont {Y.~L.}\ \bibnamefont {{L}i}}, \bibinfo {author}
  {\bibfnamefont {S.}~\bibnamefont {{C}houdhury}}, \bibinfo {author}
  {\bibfnamefont {W.}~\bibnamefont {{T}ian}}, \bibinfo {author} {\bibfnamefont
  {M.~E.}\ \bibnamefont {{H}awley}}, \bibinfo {author} {\bibfnamefont
  {B.}~\bibnamefont {{C}raigo}}, \bibinfo {author} {\bibfnamefont {A.~K.}\
  \bibnamefont {{T}agantsev}}, \bibinfo {author} {\bibfnamefont {X.~Q.}\
  \bibnamefont {{P}an}}, \bibinfo {author} {\bibfnamefont {S.~K.}\ \bibnamefont
  {{S}treiffer}}, \bibinfo {author} {\bibfnamefont {L.~Q.}\ \bibnamefont
  {{C}hen}}, \bibinfo {author} {\bibfnamefont {S.~W.}\ \bibnamefont
  {{K}irchoefer}}, \bibinfo {author} {\bibfnamefont {J.}~\bibnamefont
  {{L}evy}}, \ and\ \bibinfo {author} {\bibfnamefont {D.~G.}\ \bibnamefont
  {{S}chlom}},\ }\href {\doibase 10.1038/nature02773} {\bibfield  {journal}
  {\bibinfo  {journal} {{N}ature}\ }\textbf {\bibinfo {volume} {430}},\
  \bibinfo {pages} {758} (\bibinfo {year} {2004})}\BibitemShut {NoStop}%
\bibitem [{\citenamefont {{L}unkenheimer}\ \emph {et~al.}(2002)\citenamefont
  {{L}unkenheimer}, \citenamefont {{B}obnar}, \citenamefont {{P}ronin},
  \citenamefont {{R}itus}, \citenamefont {{V}olkov},\ and\ \citenamefont
  {{L}oidl}}]{Lunkenheimer2002}%
  \BibitemOpen
  \bibfield  {author} {\bibinfo {author} {\bibfnamefont {P.}~\bibnamefont
  {{L}unkenheimer}}, \bibinfo {author} {\bibfnamefont {V.}~\bibnamefont
  {{B}obnar}}, \bibinfo {author} {\bibfnamefont {A.~V.}\ \bibnamefont
  {{P}ronin}}, \bibinfo {author} {\bibfnamefont {A.~I.}\ \bibnamefont
  {{R}itus}}, \bibinfo {author} {\bibfnamefont {A.~A.}\ \bibnamefont
  {{V}olkov}}, \ and\ \bibinfo {author} {\bibfnamefont {A.}~\bibnamefont
  {{L}oidl}},\ }\href {\doibase 10.1103/PhysRevB.66.052105} {\bibfield
  {journal} {\bibinfo  {journal} {{P}hys. {R}ev. {B}}\ }\textbf {\bibinfo
  {volume} {66}},\ \bibinfo {pages} {052105} (\bibinfo {year}
  {2002})}\BibitemShut {NoStop}%
\bibitem [{\citenamefont {{L}angenberg}\ \emph {et~al.}(2012)\citenamefont
  {{L}angenberg}, \citenamefont {{F}ina}, \citenamefont {{V}entura},
  \citenamefont {{N}oheda}, \citenamefont {{V}arela},\ and\ \citenamefont
  {{F}ontcuberta}}]{Langenberg2012}%
  \BibitemOpen
  \bibfield  {author} {\bibinfo {author} {\bibfnamefont {E.}~\bibnamefont
  {{L}angenberg}}, \bibinfo {author} {\bibfnamefont {I.}~\bibnamefont
  {{F}ina}}, \bibinfo {author} {\bibfnamefont {J.}~\bibnamefont {{V}entura}},
  \bibinfo {author} {\bibfnamefont {B.}~\bibnamefont {{N}oheda}}, \bibinfo
  {author} {\bibfnamefont {M.}~\bibnamefont {{V}arela}}, \ and\ \bibinfo
  {author} {\bibfnamefont {J.}~\bibnamefont {{F}ontcuberta}},\ }\href {\doibase
  10.1103/PhysRevB.86.085108} {\bibfield  {journal} {\bibinfo  {journal}
  {{P}hys. {R}ev. {B}}\ }\textbf {\bibinfo {volume} {86}},\ \bibinfo {pages}
  {085108} (\bibinfo {year} {2012})}\BibitemShut {NoStop}%
\bibitem [{\citenamefont {Kalinin}\ and\ \citenamefont
  {Gruverman}(2006)}]{Kalinin2006}%
  \BibitemOpen
  \bibinfo {editor} {\bibfnamefont {S.}~\bibnamefont {Kalinin}}\ and\ \bibinfo
  {editor} {\bibfnamefont {A.}~\bibnamefont {Gruverman}},\ eds.,\ \href@noop {}
  {\emph {\bibinfo {title} {{S}canning probe microscopy: {E}lectrical and
  electrochemical phenomena at the nanoscale}}}\ (\bibinfo  {publisher}
  {Springer, New York},\ \bibinfo {year} {2006})\BibitemShut {NoStop}%
\bibitem [{\citenamefont {{S}treiffer}\ \emph {et~al.}(2002)\citenamefont
  {{S}treiffer}, \citenamefont {{E}astman}, \citenamefont {{F}ong},
  \citenamefont {{T}hompson}, \citenamefont {{M}unkholm}, \citenamefont
  {{R}amana {M}urty}, \citenamefont {{A}uciello}, \citenamefont {{B}ai},\ and\
  \citenamefont {{S}tephenson}}]{Streiffer2002}%
  \BibitemOpen
  \bibfield  {author} {\bibinfo {author} {\bibfnamefont {S.}~\bibnamefont
  {{S}treiffer}}, \bibinfo {author} {\bibfnamefont {J.}~\bibnamefont
  {{E}astman}}, \bibinfo {author} {\bibfnamefont {D.}~\bibnamefont {{F}ong}},
  \bibinfo {author} {\bibfnamefont {C.}~\bibnamefont {{T}hompson}}, \bibinfo
  {author} {\bibfnamefont {A.}~\bibnamefont {{M}unkholm}}, \bibinfo {author}
  {\bibfnamefont {M.}~\bibnamefont {{R}amana {M}urty}}, \bibinfo {author}
  {\bibfnamefont {O.}~\bibnamefont {{A}uciello}}, \bibinfo {author}
  {\bibfnamefont {G.}~\bibnamefont {{B}ai}}, \ and\ \bibinfo {author}
  {\bibfnamefont {G.}~\bibnamefont {{S}tephenson}},\ }\href {\doibase
  10.1103/PhysRevLett.89.067601} {\bibfield  {journal} {\bibinfo  {journal}
  {{P}hys. {R}ev. {L}ett.}\ }\textbf {\bibinfo {volume} {89}},\ \bibinfo
  {pages} {067601} (\bibinfo {year} {2002})}\BibitemShut {NoStop}%
\bibitem [{\citenamefont {{F}ong}\ \emph {et~al.}(2004)\citenamefont {{F}ong},
  \citenamefont {{S}tephenson}, \citenamefont {{S}treiffer}, \citenamefont
  {{E}astman}, \citenamefont {{A}uciello}, \citenamefont {{F}uoss},\ and\
  \citenamefont {{T}hompson}}]{Fong2004}%
  \BibitemOpen
  \bibfield  {author} {\bibinfo {author} {\bibfnamefont {D.~D.}\ \bibnamefont
  {{F}ong}}, \bibinfo {author} {\bibfnamefont {G.~B.}\ \bibnamefont
  {{S}tephenson}}, \bibinfo {author} {\bibfnamefont {S.~K.}\ \bibnamefont
  {{S}treiffer}}, \bibinfo {author} {\bibfnamefont {J.~A.}\ \bibnamefont
  {{E}astman}}, \bibinfo {author} {\bibfnamefont {O.}~\bibnamefont
  {{A}uciello}}, \bibinfo {author} {\bibfnamefont {P.~H.}\ \bibnamefont
  {{F}uoss}}, \ and\ \bibinfo {author} {\bibfnamefont {C.}~\bibnamefont
  {{T}hompson}},\ }\href {\doibase 10.1126/science.1098252} {\bibfield
  {journal} {\bibinfo  {journal} {{S}cience}\ }\textbf {\bibinfo {volume}
  {304}},\ \bibinfo {pages} {1650} (\bibinfo {year} {2004})}\BibitemShut
  {NoStop}%
\bibitem [{\citenamefont {{T}enne}\ \emph {et~al.}(2006)\citenamefont
  {{T}enne}, \citenamefont {{B}ruchhausen}, \citenamefont {{L}anzillotti
  {K}imura}, \citenamefont {{F}ainstein}, \citenamefont {{K}atiyar},
  \citenamefont {{C}antarero}, \citenamefont {{S}oukiassian}, \citenamefont
  {{V}aithyanathan}, \citenamefont {{H}aeni}, \citenamefont {{T}ian},
  \citenamefont {{S}chlom}, \citenamefont {{C}hoi}, \citenamefont {{K}im},
  \citenamefont {{E}om}, \citenamefont {{S}un}, \citenamefont {{P}an},
  \citenamefont {{L}i}, \citenamefont {{C}hen}, \citenamefont {{J}ia},
  \citenamefont {{N}akhmanson}, \citenamefont {{R}abe},\ and\ \citenamefont
  {{X}i}}]{Tenne2006}%
  \BibitemOpen
  \bibfield  {author} {\bibinfo {author} {\bibfnamefont {D.~A.}\ \bibnamefont
  {{T}enne}}, \bibinfo {author} {\bibfnamefont {A.}~\bibnamefont
  {{B}ruchhausen}}, \bibinfo {author} {\bibfnamefont {N.~D.}\ \bibnamefont
  {{L}anzillotti {K}imura}}, \bibinfo {author} {\bibfnamefont {A.}~\bibnamefont
  {{F}ainstein}}, \bibinfo {author} {\bibfnamefont {R.~S.}\ \bibnamefont
  {{K}atiyar}}, \bibinfo {author} {\bibfnamefont {A.}~\bibnamefont
  {{C}antarero}}, \bibinfo {author} {\bibfnamefont {A.}~\bibnamefont
  {{S}oukiassian}}, \bibinfo {author} {\bibfnamefont {V.}~\bibnamefont
  {{V}aithyanathan}}, \bibinfo {author} {\bibfnamefont {J.~H.}\ \bibnamefont
  {{H}aeni}}, \bibinfo {author} {\bibfnamefont {W.}~\bibnamefont {{T}ian}},
  \bibinfo {author} {\bibfnamefont {D.~G.}\ \bibnamefont {{S}chlom}}, \bibinfo
  {author} {\bibfnamefont {K.~J.}\ \bibnamefont {{C}hoi}}, \bibinfo {author}
  {\bibfnamefont {D.~M.}\ \bibnamefont {{K}im}}, \bibinfo {author}
  {\bibfnamefont {C.~B.}\ \bibnamefont {{E}om}}, \bibinfo {author}
  {\bibfnamefont {H.~P.}\ \bibnamefont {{S}un}}, \bibinfo {author}
  {\bibfnamefont {X.~Q.}\ \bibnamefont {{P}an}}, \bibinfo {author}
  {\bibfnamefont {Y.~L.}\ \bibnamefont {{L}i}}, \bibinfo {author}
  {\bibfnamefont {L.~Q.}\ \bibnamefont {{C}hen}}, \bibinfo {author}
  {\bibfnamefont {Q.~X.}\ \bibnamefont {{J}ia}}, \bibinfo {author}
  {\bibfnamefont {S.~M.}\ \bibnamefont {{N}akhmanson}}, \bibinfo {author}
  {\bibfnamefont {K.~M.}\ \bibnamefont {{R}abe}}, \ and\ \bibinfo {author}
  {\bibfnamefont {X.~X.}\ \bibnamefont {{X}i}},\ }\href {\doibase
  10.1126/science.1130306} {\bibfield  {journal} {\bibinfo  {journal}
  {{S}cience}\ }\textbf {\bibinfo {volume} {313}},\ \bibinfo {pages} {1614}
  (\bibinfo {year} {2006})},
  \BibitemShut
  {NoStop}%
\bibitem [{Cry()}]{Crystec}%
  \BibitemOpen
  \bibinfo {howpublished} {\url{http://www.crystec.de/}}\BibitemShut {NoStop}%
\bibitem [{Sur()}]{SurfaceNet}%
  \BibitemOpen
  \bibinfo {howpublished} {\url{http://www.surfacenet.de/}}\BibitemShut {NoStop}%
\bibitem [{\citenamefont {{W}ang}\ and\ \citenamefont
  {{I}toh}(2001)}]{Wang2001}%
  \BibitemOpen
  \bibfield  {author} {\bibinfo {author} {\bibfnamefont {R.}~\bibnamefont
  {{W}ang}}\ and\ \bibinfo {author} {\bibfnamefont {M.}~\bibnamefont
  {{I}toh}},\ }\href {\doibase 10.1103/PhysRevB.64.174104} {\bibfield
  {journal} {\bibinfo  {journal} {{P}hys. {R}ev. {B}}\ }\textbf {\bibinfo
  {volume} {64}},\ \bibinfo {pages} {174104} (\bibinfo {year}
  {2001})}\BibitemShut {NoStop}%
\bibitem [{Woo()}]{Woollam}%
  \BibitemOpen
  \bibinfo {howpublished} {\url{http://www.jawoollam.com/}}\BibitemShut
  {NoStop}%
\bibitem [{\citenamefont {{B}ernhard}\ \emph {et~al.}(2004)\citenamefont
  {{B}ernhard}, \citenamefont {{H}uml\'{i}\v{c}ek},\ and\ \citenamefont
  {{K}eimer}}]{Bernhard2004}%
  \BibitemOpen
  \bibfield  {author} {\bibinfo {author} {\bibfnamefont {C.}~\bibnamefont
  {{B}ernhard}}, \bibinfo {author} {\bibfnamefont {J.}~\bibnamefont
  {{H}uml\'{i}\v{c}ek}}, \ and\ \bibinfo {author} {\bibfnamefont
  {B.}~\bibnamefont {{K}eimer}},\ }\href {\doibase 10.1016/j.tsf.2004.01.002}
  {\bibfield  {journal} {\bibinfo  {journal} {{T}hin {S}olid {F}ilms}\ }\textbf
  {\bibinfo {volume} {455-456}},\ \bibinfo {pages} {143} (\bibinfo {year}
  {2004})}\BibitemShut {NoStop}%
\bibitem [{\citenamefont {{K}im}\ \emph {et~al.}(2010)\citenamefont {{K}im},
  \citenamefont {{R}\"ossle}, \citenamefont {{D}ubroka}, \citenamefont
  {{M}alik}, \citenamefont {{W}olf},\ and\ \citenamefont
  {{B}ernhard}}]{Kim2010}%
  \BibitemOpen
  \bibfield  {author} {\bibinfo {author} {\bibfnamefont {K.~W.}\ \bibnamefont
  {{K}im}}, \bibinfo {author} {\bibfnamefont {M.}~\bibnamefont {{R}\"ossle}},
  \bibinfo {author} {\bibfnamefont {A.}~\bibnamefont {{D}ubroka}}, \bibinfo
  {author} {\bibfnamefont {V.~K.}\ \bibnamefont {{M}alik}}, \bibinfo {author}
  {\bibfnamefont {T.}~\bibnamefont {{W}olf}}, \ and\ \bibinfo {author}
  {\bibfnamefont {C.}~\bibnamefont {{B}ernhard}},\ }\href {\doibase
  10.1103/PhysRevB.81.214508} {\bibfield  {journal} {\bibinfo  {journal}
  {{P}hys. {R}ev. {B}}\ }\textbf {\bibinfo {volume} {81}},\ \bibinfo {pages}
  {214508} (\bibinfo {year} {2010})}\BibitemShut {NoStop}%
\bibitem [{\citenamefont {{C}ardona}(1965)}]{Cardona1965}%
  \BibitemOpen
  \bibfield  {author} {\bibinfo {author} {\bibfnamefont {M.}~\bibnamefont
  {{C}ardona}},\ }\href@noop {} {\bibfield  {journal} {\bibinfo  {journal}
  {{P}hys. {R}ev.}\ }\textbf {\bibinfo {volume} {140}},\  (\bibinfo {year}
  {1965})}\BibitemShut {NoStop}%
\bibitem [{\citenamefont {van {B}enthem}\ \emph {et~al.}(2001)\citenamefont
  {van {B}enthem}, \citenamefont {{E}lsaesser},\ and\ \citenamefont
  {{F}rench}}]{Benthem2001}%
  \BibitemOpen
  \bibfield  {author} {\bibinfo {author} {\bibfnamefont {K.}~\bibnamefont {van
  {B}enthem}}, \bibinfo {author} {\bibfnamefont {C.}~\bibnamefont
  {{E}lsaesser}}, \ and\ \bibinfo {author} {\bibfnamefont {R.~H.}\ \bibnamefont
  {{F}rench}},\ }\href {\doibase 10.1063/1.1415766͔} {\bibfield  {journal}
  {\bibinfo  {journal} {{J}. {A}ppl. {P}hys.}\ }\textbf {\bibinfo {volume}
  {90}},\ \bibinfo {pages} {6156} (\bibinfo {year} {2001})}\BibitemShut
  {NoStop}%
\bibitem [{\citenamefont {{S}avitzky}\ and\ \citenamefont
  {{G}olay}(1964)}]{Savitzky1964}%
  \BibitemOpen
  \bibfield  {author} {\bibinfo {author} {\bibfnamefont {A.}~\bibnamefont
  {{S}avitzky}}\ and\ \bibinfo {author} {\bibfnamefont {M.~J.~E.}\ \bibnamefont
  {{G}olay}},\ }\href {\doibase 10.1021/ac60214a047} {\bibfield  {journal}
  {\bibinfo  {journal} {{A}nal. {C}hem.}\ }\textbf {\bibinfo {volume} {36}},\
  \bibinfo {pages} {1627} (\bibinfo {year} {1964})}
  \BibitemShut {NoStop}%
\bibitem [{\citenamefont {{Y}u}\ and\ \citenamefont
  {{C}ardona}(2005)}]{Yu2005}%
  \BibitemOpen
  \bibfield  {author} {\bibinfo {author} {\bibfnamefont {P.~Y.}\ \bibnamefont
  {{Y}u}}\ and\ \bibinfo {author} {\bibfnamefont {M.}~\bibnamefont
  {{C}ardona}},\ }\href@noop {} {\emph {\bibinfo {title} {{F}undamentals of
  {S}emiconductors}}},\ \bibinfo {edition} {3rd}\ ed.\ (\bibinfo  {publisher}
  {Springer Verlag Berlin},\ \bibinfo {year} {2005})\BibitemShut {NoStop}%
\bibitem [{\citenamefont {{L}autenschlager}\ \emph {et~al.}(1987)\citenamefont
  {{L}autenschlager}, \citenamefont {{G}arriga}, \citenamefont
  {{L}ogothetidis},\ and\ \citenamefont {{C}ardona}}]{Lautenschlager1987}%
  \BibitemOpen
  \bibfield  {author} {\bibinfo {author} {\bibfnamefont {P.}~\bibnamefont
  {{L}autenschlager}}, \bibinfo {author} {\bibfnamefont {M.}~\bibnamefont
  {{G}arriga}}, \bibinfo {author} {\bibfnamefont {S.}~\bibnamefont
  {{L}ogothetidis}}, \ and\ \bibinfo {author} {\bibfnamefont {M.}~\bibnamefont
  {{C}ardona}},\ }\href {\doibase 10.1103/PhysRevB.35.9174} {\bibfield
  {journal} {\bibinfo  {journal} {{P}hys. {R}ev. {B}}\ }\textbf {\bibinfo
  {volume} {35}},\ \bibinfo {pages} {9174} (\bibinfo {year}
  {1987})}\BibitemShut {NoStop}%
\bibitem [{\citenamefont {{T}repakov}\ \emph {et~al.}(2009)\citenamefont
  {{T}repakov}, \citenamefont {{D}ejneka}, \citenamefont {{M}arkovin},
  \citenamefont {{L}ynnyk},\ and\ \citenamefont {{J}astrabik}}]{Trepakov2009}%
  \BibitemOpen
  \bibfield  {author} {\bibinfo {author} {\bibfnamefont {V.}~\bibnamefont
  {{T}repakov}}, \bibinfo {author} {\bibfnamefont {A.}~\bibnamefont
  {{D}ejneka}}, \bibinfo {author} {\bibfnamefont {P.}~\bibnamefont
  {{M}arkovin}}, \bibinfo {author} {\bibfnamefont {A.}~\bibnamefont
  {{L}ynnyk}}, \ and\ \bibinfo {author} {\bibfnamefont {L.}~\bibnamefont
  {{J}astrabik}},\ }\href {\doibase 10.1088/1367-2630/11/8/083024} {\bibfield
  {journal} {\bibinfo  {journal} {{N}ew {J}. {P}hys.}\ }\textbf {\bibinfo
  {volume} {11}},\ \bibinfo {pages} {083024} (\bibinfo {year}
  {2009})}\BibitemShut {NoStop}%
\bibitem [{\citenamefont {{L}autenschlager}\ \emph {et~al.}(1985)\citenamefont
  {{L}autenschlager}, \citenamefont {{A}llen},\ and\ \citenamefont
  {{C}ardona}}]{Lautenschlager1985}%
  \BibitemOpen
  \bibfield  {author} {\bibinfo {author} {\bibfnamefont {P.}~\bibnamefont
  {{L}autenschlager}}, \bibinfo {author} {\bibfnamefont {P.~B.}\ \bibnamefont
  {{A}llen}}, \ and\ \bibinfo {author} {\bibfnamefont {M.}~\bibnamefont
  {{C}ardona}},\ }\href {\doibase 10.1103/PhysRevB.31.2163} {\bibfield
  {journal} {\bibinfo  {journal} {{P}hys. {R}ev. {B}}\ }\textbf {\bibinfo
  {volume} {31}},\ \bibinfo {pages} {2163} (\bibinfo {year}
  {1985})}\BibitemShut {NoStop}%
\bibitem [{\citenamefont {{O}lgu\'in}\ \emph {et~al.}(2002)\citenamefont
  {{O}lgu\'in}, \citenamefont {{C}ardona},\ and\ \citenamefont
  {{C}antarero}}]{Olguin2002}%
  \BibitemOpen
  \bibfield  {author} {\bibinfo {author} {\bibfnamefont {D.}~\bibnamefont
  {{O}lgu\'in}}, \bibinfo {author} {\bibfnamefont {M.}~\bibnamefont
  {{C}ardona}}, \ and\ \bibinfo {author} {\bibfnamefont {A.}~\bibnamefont
  {{C}antarero}},\ }\href {\doibase 10.1016/S0038-1098(02)00225-9} {\bibfield
  {journal} {\bibinfo  {journal} {{S}olid {S}tate {C}ommun.}\ }\textbf
  {\bibinfo {volume} {122}},\ \bibinfo {pages} {575 } (\bibinfo {year}
  {2002})}\BibitemShut {NoStop}%
\bibitem [{\citenamefont {{K}effer}\ \emph {et~al.}(1968)\citenamefont
  {{K}effer}, \citenamefont {{H}ayes},\ and\ \citenamefont
  {{B}ienenstock}}]{Keffer1968}%
  \BibitemOpen
  \bibfield  {author} {\bibinfo {author} {\bibfnamefont {C.}~\bibnamefont
  {{K}effer}}, \bibinfo {author} {\bibfnamefont {T.~M.}\ \bibnamefont
  {{H}ayes}}, \ and\ \bibinfo {author} {\bibfnamefont {A.}~\bibnamefont
  {{B}ienenstock}},\ }\href {\doibase 10.1103/PhysRevLett.21.1676} {\bibfield
  {journal} {\bibinfo  {journal} {{P}hys. {R}ev. {L}ett.}\ }\textbf {\bibinfo
  {volume} {21}},\ \bibinfo {pages} {1676} (\bibinfo {year}
  {1968})}\BibitemShut {NoStop}%
\bibitem [{\citenamefont {{A}llen}\ and\ \citenamefont
  {{C}ardona}(1981)}]{Allen1981}%
  \BibitemOpen
  \bibfield  {author} {\bibinfo {author} {\bibfnamefont {P.~B.}\ \bibnamefont
  {{A}llen}}\ and\ \bibinfo {author} {\bibfnamefont {M.}~\bibnamefont
  {{C}ardona}},\ }\href {\doibase 10.1103/PhysRevB.23.1495} {\bibfield
  {journal} {\bibinfo  {journal} {{P}hys. {R}ev. {B}}\ }\textbf {\bibinfo
  {volume} {23}},\ \bibinfo {pages} {1495} (\bibinfo {year}
  {1981})}\BibitemShut {NoStop}%
\bibitem [{\citenamefont {{A}llen}\ and\ \citenamefont
  {{C}ardona}(1983)}]{Allen1983}%
  \BibitemOpen
  \bibfield  {author} {\bibinfo {author} {\bibfnamefont {P.~B.}\ \bibnamefont
  {{A}llen}}\ and\ \bibinfo {author} {\bibfnamefont {M.}~\bibnamefont
  {{C}ardona}},\ }\href {\doibase 10.1103/PhysRevB.27.4760} {\bibfield
  {journal} {\bibinfo  {journal} {{P}hys. {R}ev. {B}}\ }\textbf {\bibinfo
  {volume} {27}},\ \bibinfo {pages} {4760} (\bibinfo {year}
  {1983})}\BibitemShut {NoStop}%
\bibitem [{\citenamefont {{C}ardona}(2005)}]{Cardona2005a}%
  \BibitemOpen
  \bibfield  {author} {\bibinfo {author} {\bibfnamefont {M.}~\bibnamefont
  {{C}ardona}},\ }\href
  {http://www.sciencedirect.com/science/article/pii/S0038109804009160}
  {\bibfield  {journal} {\bibinfo  {journal} {{S}olid {S}tate {C}ommun.}\
  }\textbf {\bibinfo {volume} {133}},\ \bibinfo {pages} {3} (\bibinfo {year}
  {2005})}\BibitemShut {NoStop}%
\bibitem [{\citenamefont {{B}hosale}\ \emph {et~al.}(2012)\citenamefont
  {{B}hosale}, \citenamefont {{R}amdas}, \citenamefont {{B}urger},
  \citenamefont {{M}u{\~n}oz}, \citenamefont {{R}omero}, \citenamefont
  {{C}ardona}, \citenamefont {{L}auck},\ and\ \citenamefont
  {{K}remer}}]{Bhosale2012}%
  \BibitemOpen
  \bibfield  {author} {\bibinfo {author} {\bibfnamefont {J.}~\bibnamefont
  {{B}hosale}}, \bibinfo {author} {\bibfnamefont {A.~K.}\ \bibnamefont
  {{R}amdas}}, \bibinfo {author} {\bibfnamefont {A.}~\bibnamefont {{B}urger}},
  \bibinfo {author} {\bibfnamefont {A.}~\bibnamefont {{M}u{\~n}oz}}, \bibinfo
  {author} {\bibfnamefont {A.~H.}\ \bibnamefont {{R}omero}}, \bibinfo {author}
  {\bibfnamefont {M.}~\bibnamefont {{C}ardona}}, \bibinfo {author}
  {\bibfnamefont {R.}~\bibnamefont {{L}auck}}, \ and\ \bibinfo {author}
  {\bibfnamefont {R.~K.}\ \bibnamefont {{K}remer}},\ }\href {\doibase
  10.1103/PhysRevB.86.195208} {\bibfield  {journal} {\bibinfo  {journal}
  {{P}hys. {R}ev. {B}}\ }\textbf {\bibinfo {volume} {86}},\ \bibinfo {pages}
  {195208} (\bibinfo {year} {2012})}\BibitemShut {NoStop}%
\bibitem [{\citenamefont {{F}an}(1951)}]{Fan1951}%
  \BibitemOpen
  \bibfield  {author} {\bibinfo {author} {\bibfnamefont {H.~Y.}\ \bibnamefont
  {{F}an}},\ }\href {\doibase 10.1103/PhysRev.82.900} {\bibfield  {journal}
  {\bibinfo  {journal} {{P}hys. {R}ev.}\ }\textbf {\bibinfo {volume} {82}},\
  \bibinfo {pages} {900} (\bibinfo {year} {1951})}\BibitemShut {NoStop}%
\bibitem [{\citenamefont {{L}iu}\ \emph {et~al.}(1997)\citenamefont {{L}iu},
  \citenamefont {{F}inlayson},\ and\ \citenamefont {{S}mith}}]{Liu1997}%
  \BibitemOpen
  \bibfield  {author} {\bibinfo {author} {\bibfnamefont {M.}~\bibnamefont
  {{L}iu}}, \bibinfo {author} {\bibfnamefont {T.~R.}\ \bibnamefont
  {{F}inlayson}}, \ and\ \bibinfo {author} {\bibfnamefont {T.~F.}\ \bibnamefont
  {{S}mith}},\ }\href {\doibase 10.1103/PhysRevB.55.3480} {\bibfield  {journal}
  {\bibinfo  {journal} {{P}hys. {R}ev. {B}}\ }\textbf {\bibinfo {volume}
  {55}},\ \bibinfo {pages} {3480} (\bibinfo {year} {1997})}\BibitemShut
  {NoStop}%
\bibitem [{\citenamefont {{L}oetzsch}\ \emph {et~al.}(2010)\citenamefont
  {{L}oetzsch}, \citenamefont {{L}\"ubcke}, \citenamefont {{U}schmann},
  \citenamefont {{F}orster}, \citenamefont {{G}rosse}, \citenamefont
  {{T}huerk}, \citenamefont {{K}oettig}, \citenamefont {{S}chmidl},\ and\
  \citenamefont {{S}eidel}}]{Loetzsch2010}%
  \BibitemOpen
  \bibfield  {author} {\bibinfo {author} {\bibfnamefont {R.}~\bibnamefont
  {{L}oetzsch}}, \bibinfo {author} {\bibfnamefont {A.}~\bibnamefont
  {{L}\"ubcke}}, \bibinfo {author} {\bibfnamefont {I.}~\bibnamefont
  {{U}schmann}}, \bibinfo {author} {\bibfnamefont {E.}~\bibnamefont
  {{F}orster}}, \bibinfo {author} {\bibfnamefont {V.}~\bibnamefont {{G}rosse}},
  \bibinfo {author} {\bibfnamefont {M.}~\bibnamefont {{T}huerk}}, \bibinfo
  {author} {\bibfnamefont {T.}~\bibnamefont {{K}oettig}}, \bibinfo {author}
  {\bibfnamefont {F.}~\bibnamefont {{S}chmidl}}, \ and\ \bibinfo {author}
  {\bibfnamefont {P.}~\bibnamefont {{S}eidel}},\ }\href {\doibase
  10.1063/1.3324695} {\bibfield  {journal} {\bibinfo  {journal} {{A}ppl.
  {P}hys. {L}ett.}\ }\textbf {\bibinfo {volume} {96}},\ \bibinfo {eid} {071901}
  (\bibinfo {year} {2010})}\BibitemShut {NoStop}%
\bibitem [{\citenamefont {{G}hebouli}\ \emph {et~al.}(2009)\citenamefont
  {{G}hebouli}, \citenamefont {{G}hebouli}, \citenamefont {{C}hihi},
  \citenamefont {{F}atmi}, \citenamefont {{B}oucetta},\ and\ \citenamefont
  {{R}effas}}]{Ghebouli2009}%
  \BibitemOpen
  \bibfield  {author} {\bibinfo {author} {\bibfnamefont {B.}~\bibnamefont
  {{G}hebouli}}, \bibinfo {author} {\bibfnamefont {M.}~\bibnamefont
  {{G}hebouli}}, \bibinfo {author} {\bibfnamefont {T.}~\bibnamefont {{C}hihi}},
  \bibinfo {author} {\bibfnamefont {M.}~\bibnamefont {{F}atmi}}, \bibinfo
  {author} {\bibfnamefont {S.}~\bibnamefont {{B}oucetta}}, \ and\ \bibinfo
  {author} {\bibfnamefont {M.}~\bibnamefont {{R}effas}},\ }\href {\doibase
  10.1016/j.ssc.2009.09.001} {\bibfield  {journal} {\bibinfo  {journal}
  {{S}olid {S}tate {C}ommun.}\ }\textbf {\bibinfo {volume} {149}},\ \bibinfo
  {pages} {2244 } (\bibinfo {year} {2009})}\BibitemShut {NoStop}%
\bibitem [{\citenamefont {{L}ytle}(1964)}]{Lytle1964}%
  \BibitemOpen
  \bibfield  {author} {\bibinfo {author} {\bibfnamefont {F.~W.}\ \bibnamefont
  {{L}ytle}},\ }\href@noop {} {\bibfield  {journal} {\bibinfo  {journal} {{J}.
  {A}ppl. {P}hys.}\ }\textbf {\bibinfo {volume} {35}},\ \bibinfo {pages} {2212}
  (\bibinfo {year} {1964})}\BibitemShut {NoStop}%
\bibitem [{\citenamefont {{M}\"{u}ller}\ and\ \citenamefont
  {{B}urkard}(1979)}]{Muller1979}%
  \BibitemOpen
  \bibfield  {author} {\bibinfo {author} {\bibfnamefont {K.~A.}\ \bibnamefont
  {{M}\"{u}ller}}\ and\ \bibinfo {author} {\bibfnamefont {H.}~\bibnamefont
  {{B}urkard}},\ }\href@noop {} {\bibfield  {journal} {\bibinfo  {journal}
  {{P}hys. {R}ev. {B}}\ }\textbf {\bibinfo {volume} {19}},\ \bibinfo {pages}
  {3593} (\bibinfo {year} {1979})}\BibitemShut {NoStop}%
\bibitem [{\citenamefont {{V}ogt}(1995)}]{Vogt1995}%
  \BibitemOpen
  \bibfield  {author} {\bibinfo {author} {\bibfnamefont {H.}~\bibnamefont
  {{V}ogt}},\ }\href {\doibase 10.1103/PhysRevB.51.8046} {\bibfield  {journal}
  {\bibinfo  {journal} {{P}hys. {R}ev. {B}}\ }\textbf {\bibinfo {volume}
  {51}},\ \bibinfo {pages} {8046} (\bibinfo {year} {1995})}\BibitemShut
  {NoStop}%
\bibitem [{\citenamefont {{S}ervoin}\ \emph {et~al.}(1980)\citenamefont
  {{S}ervoin}, \citenamefont {{L}uspin},\ and\ \citenamefont
  {{G}ervais}}]{Servoin1980}%
  \BibitemOpen
  \bibfield  {author} {\bibinfo {author} {\bibfnamefont {J.~L.}\ \bibnamefont
  {{S}ervoin}}, \bibinfo {author} {\bibfnamefont {Y.}~\bibnamefont {{L}uspin}},
  \ and\ \bibinfo {author} {\bibfnamefont {F.}~\bibnamefont {{G}ervais}},\
  }\href@noop {} {\bibfield  {journal} {\bibinfo  {journal} {{P}hys. {R}ev.
  {B}}\ }\textbf {\bibinfo {volume} {22}},\ \bibinfo {pages} {5501} (\bibinfo
  {year} {1980})}\BibitemShut {NoStop}%
\bibitem [{\citenamefont {{S}irenko}\ \emph {et~al.}(2000)\citenamefont
  {{S}irenko}, \citenamefont {{B}ernhard}, \citenamefont {{G}olnik},
  \citenamefont {{C}lark}, \citenamefont {{H}ao}, \citenamefont {{S}i},\ and\
  \citenamefont {{X}i}}]{Sirenko2000}%
  \BibitemOpen
  \bibfield  {author} {\bibinfo {author} {\bibfnamefont {A.~A.}\ \bibnamefont
  {{S}irenko}}, \bibinfo {author} {\bibfnamefont {C.}~\bibnamefont
  {{B}ernhard}}, \bibinfo {author} {\bibfnamefont {A.}~\bibnamefont
  {{G}olnik}}, \bibinfo {author} {\bibfnamefont {A.~M.}\ \bibnamefont
  {{C}lark}}, \bibinfo {author} {\bibfnamefont {J.}~\bibnamefont {{H}ao}},
  \bibinfo {author} {\bibfnamefont {W.}~\bibnamefont {{S}i}}, \ and\ \bibinfo
  {author} {\bibfnamefont {X.~X.}\ \bibnamefont {{X}i}},\ }\href@noop {}
  {\bibfield  {journal} {\bibinfo  {journal} {{N}ature}\ }\textbf {\bibinfo
  {volume} {404}},\ \bibinfo {pages} {373} (\bibinfo {year}
  {2000})}\BibitemShut {NoStop}%
\bibitem [{SOM()}]{SOM}%
  \BibitemOpen
  \href@noop {} {\enquote {\bibinfo {title} {{S}ee {S}upplemental {M}aterial at
  [{URL} will be inserted by publisher] for typical ellipsometry spectra of perovskite materials, the description of the fitting procedure of the direct interband transition, and a detailed discussion of the temperature behavior of $\omega_{\mathrm{TO}}$ of the investigated perovskites and the $T$-dependence of $\omega_{\mathrm{LO}}$ and $\varepsilon_{\infty}$.}}\ }\BibitemShut {NoStop}%
\bibitem [{\citenamefont {{P}erry}\ and\ \citenamefont
  {{M}c{N}elly}(1967)}]{Perry1967}%
  \BibitemOpen
  \bibfield  {author} {\bibinfo {author} {\bibfnamefont {C.~H.}\ \bibnamefont
  {{P}erry}}\ and\ \bibinfo {author} {\bibfnamefont {T.~F.}\ \bibnamefont
  {{M}c{N}elly}},\ }\href@noop {} {\bibfield  {journal} {\bibinfo  {journal}
  {{P}hys. {R}ev.}\ }\textbf {\bibinfo {volume} {154}},\ \bibinfo {pages} {456}
  (\bibinfo {year} {1967})}\BibitemShut {NoStop}%
\bibitem [{\citenamefont {{F}leury}\ and\ \citenamefont
  {{W}orlock}(1968)}]{Fleury1968}%
  \BibitemOpen
  \bibfield  {author} {\bibinfo {author} {\bibfnamefont {P.~A.}\ \bibnamefont
  {{F}leury}}\ and\ \bibinfo {author} {\bibfnamefont {J.~M.}\ \bibnamefont
  {{W}orlock}},\ }\href@noop {} {\bibfield  {journal} {\bibinfo  {journal}
  {{P}hys. {R}ev.}\ }\textbf {\bibinfo {volume} {174}},\ \bibinfo {pages} {613}
  (\bibinfo {year} {1968})}\BibitemShut {NoStop}%
\bibitem [{\citenamefont {{I}toh}\ \emph {et~al.}(1999)\citenamefont {{I}toh},
  \citenamefont {{W}ang}, \citenamefont {{I}naguma}, \citenamefont
  {{Y}amaguchi}, \citenamefont {{S}han},\ and\ \citenamefont
  {{N}akamura}}]{Itoh1999}%
  \BibitemOpen
  \bibfield  {author} {\bibinfo {author} {\bibfnamefont {M.}~\bibnamefont
  {{I}toh}}, \bibinfo {author} {\bibfnamefont {R.}~\bibnamefont {{W}ang}},
  \bibinfo {author} {\bibfnamefont {Y.}~\bibnamefont {{I}naguma}}, \bibinfo
  {author} {\bibfnamefont {T.}~\bibnamefont {{Y}amaguchi}}, \bibinfo {author}
  {\bibfnamefont {Y.-J.}\ \bibnamefont {{S}han}}, \ and\ \bibinfo {author}
  {\bibfnamefont {T.}~\bibnamefont {{N}akamura}},\ }\href {\doibase
  10.1103/PhysRevLett.82.3540} {\bibfield  {journal} {\bibinfo  {journal}
  {{P}hys. {R}ev. {L}ett.}\ }\textbf {\bibinfo {volume} {82}},\ \bibinfo
  {pages} {3540} (\bibinfo {year} {1999})}\BibitemShut {NoStop}%
\bibitem [{\citenamefont {{B}erger}\ \emph {et~al.}(2011)\citenamefont
  {{B}erger}, \citenamefont {{F}ennie},\ and\ \citenamefont
  {{N}eaton}}]{Berger2011}%
  \BibitemOpen
  \bibfield  {author} {\bibinfo {author} {\bibfnamefont {R.~F.}\ \bibnamefont
  {{B}erger}}, \bibinfo {author} {\bibfnamefont {C.~J.}\ \bibnamefont
  {{F}ennie}}, \ and\ \bibinfo {author} {\bibfnamefont {J.~B.}\ \bibnamefont
  {{N}eaton}},\ }\href {\doibase 10.1103/PhysRevLett.107.146804} {\bibfield
  {journal} {\bibinfo  {journal} {{P}hys. {R}ev. {L}ett.}\ }\textbf {\bibinfo
  {volume} {107}},\ \bibinfo {pages} {146804} (\bibinfo {year}
  {2011})}\BibitemShut {NoStop}%
\bibitem [{\citenamefont {{H}linka}\ \emph {et~al.}(2008)\citenamefont
  {{H}linka}, \citenamefont {{O}stapchuk}, \citenamefont {{N}uzhnyy},
  \citenamefont {{P}etzelt}, \citenamefont {{K}uzel}, \citenamefont {{K}adlec},
  \citenamefont {{V}anek}, \citenamefont {{P}onomareva},\ and\ \citenamefont
  {{B}ellaiche}}]{Hlinka2008}%
  \BibitemOpen
  \bibfield  {author} {\bibinfo {author} {\bibfnamefont {J.}~\bibnamefont
  {{H}linka}}, \bibinfo {author} {\bibfnamefont {T.}~\bibnamefont
  {{O}stapchuk}}, \bibinfo {author} {\bibfnamefont {D.}~\bibnamefont
  {{N}uzhnyy}}, \bibinfo {author} {\bibfnamefont {J.}~\bibnamefont
  {{P}etzelt}}, \bibinfo {author} {\bibfnamefont {P.}~\bibnamefont {{K}uzel}},
  \bibinfo {author} {\bibfnamefont {C.}~\bibnamefont {{K}adlec}}, \bibinfo
  {author} {\bibfnamefont {P.}~\bibnamefont {{V}anek}}, \bibinfo {author}
  {\bibfnamefont {I.}~\bibnamefont {{P}onomareva}}, \ and\ \bibinfo {author}
  {\bibfnamefont {L.}~\bibnamefont {{B}ellaiche}},\ }\href {\doibase
  10.1103/PhysRevLett.101.167402} {\bibfield  {journal} {\bibinfo  {journal}
  {{P}hys. {R}ev. {L}ett.}\ }\textbf {\bibinfo {volume} {101}},\ \bibinfo
  {pages} {167402} (\bibinfo {year} {2008})}\BibitemShut {NoStop}%
\bibitem [{\citenamefont {{P}onomareva}\ \emph {et~al.}(2008)\citenamefont
  {{P}onomareva}, \citenamefont {{B}ellaiche}, \citenamefont {{O}stapchuk},
  \citenamefont {{H}linka},\ and\ \citenamefont {{P}etzelt}}]{Ponomareva2008}%
  \BibitemOpen
  \bibfield  {author} {\bibinfo {author} {\bibfnamefont {I.}~\bibnamefont
  {{P}onomareva}}, \bibinfo {author} {\bibfnamefont {L.}~\bibnamefont
  {{B}ellaiche}}, \bibinfo {author} {\bibfnamefont {T.}~\bibnamefont
  {{O}stapchuk}}, \bibinfo {author} {\bibfnamefont {J.}~\bibnamefont
  {{H}linka}}, \ and\ \bibinfo {author} {\bibfnamefont {J.}~\bibnamefont
  {{P}etzelt}},\ }\href {\doibase 10.1103/PhysRevB.77.012102} {\bibfield
  {journal} {\bibinfo  {journal} {{P}hys. {R}ev. {B}}\ }\textbf {\bibinfo
  {volume} {77}},\ \bibinfo {pages} {012102} (\bibinfo {year}
  {2008})}\BibitemShut {NoStop}%
\bibitem [{\citenamefont {{V}ogt}\ \emph {et~al.}(1982)\citenamefont {{V}ogt},
  \citenamefont {{S}anjurjo},\ and\ \citenamefont {{R}ossbroich}}]{Vogt1982}%
  \BibitemOpen
  \bibfield  {author} {\bibinfo {author} {\bibfnamefont {H.}~\bibnamefont
  {{V}ogt}}, \bibinfo {author} {\bibfnamefont {J.~A.}\ \bibnamefont
  {{S}anjurjo}}, \ and\ \bibinfo {author} {\bibfnamefont {G.}~\bibnamefont
  {{R}ossbroich}},\ }\href {\doibase 10.1103/PhysRevB.26.5904} {\bibfield
  {journal} {\bibinfo  {journal} {{P}hys. {R}ev. {B}}\ }\textbf {\bibinfo
  {volume} {26}},\ \bibinfo {pages} {5904} (\bibinfo {year}
  {1982})}\BibitemShut {NoStop}%
\bibitem [{\citenamefont {{N}uzhnyy}\ \emph {et~al.}(2009)\citenamefont
  {{N}uzhnyy}, \citenamefont {{P}etzelt}, \citenamefont {{K}amba},
  \citenamefont {{K}uzel}, \citenamefont {{K}adlec}, \citenamefont {{B}ovtun},
  \citenamefont {{K}empa}, \citenamefont {{S}chubert}, \citenamefont
  {{B}rooks},\ and\ \citenamefont {{S}chlom}}]{Nuzhnyy2009}%
  \BibitemOpen
  \bibfield  {author} {\bibinfo {author} {\bibfnamefont {D.}~\bibnamefont
  {{N}uzhnyy}}, \bibinfo {author} {\bibfnamefont {J.}~\bibnamefont
  {{P}etzelt}}, \bibinfo {author} {\bibfnamefont {S.}~\bibnamefont {{K}amba}},
  \bibinfo {author} {\bibfnamefont {P.}~\bibnamefont {{K}uzel}}, \bibinfo
  {author} {\bibfnamefont {C.}~\bibnamefont {{K}adlec}}, \bibinfo {author}
  {\bibfnamefont {V.}~\bibnamefont {{B}ovtun}}, \bibinfo {author}
  {\bibfnamefont {M.}~\bibnamefont {{K}empa}}, \bibinfo {author} {\bibfnamefont
  {J.}~\bibnamefont {{S}chubert}}, \bibinfo {author} {\bibfnamefont {C.~M.}\
  \bibnamefont {{B}rooks}}, \ and\ \bibinfo {author} {\bibfnamefont {D.~G.}\
  \bibnamefont {{S}chlom}},\ }\href@noop {} {\bibfield  {journal} {\bibinfo
  {journal} {{A}ppl. {P}hys. {L}ett.}\ }\textbf {\bibinfo {volume} {95}},\
  \bibinfo {pages} {232902} (\bibinfo {year} {2009})}\BibitemShut {NoStop}%
\end{thebibliography}
%


\begin{widetext}


\section*{Supplementary material}

# Optical probe of ferroelectric order in bulk and thin film perovskite titanates

## Ellipsometry spectra of BaTiO$_3$, KTaO$_3$, and SrTi$^{18}$O$_3$ in the NIR-UV

In figure S1 we show typical ellipsometry spectra for the incipient ferroelectric material KTaO$_3$ and for CaTiO$_3$ in terms of the real and imaginary parts of the dielectric function, $\varepsilon = \varepsilon_1 + i\varepsilon_2$, together with the corresponding second derivative spectra of the imaginary part, $\frac{\partial^2 \varepsilon_2}{\partial E^2}$. Corresponding data for the ferroelectric materials SrTi$^{18}$O$_3$ and BaTiO$_3$ are shown in figure S2. All presented spectra have been corrected for surface roughness effects using the Woollam VASE software [1]. The surface roughness has been found to be on the order of $\leq 4$ nm for all investigated samples and it showed only a weak variation with temperature during a measurement cycle. Only during the measurements at $T > 550$ K we have sometimes observed a decrease in the roughness that occurred during or after the measurements.

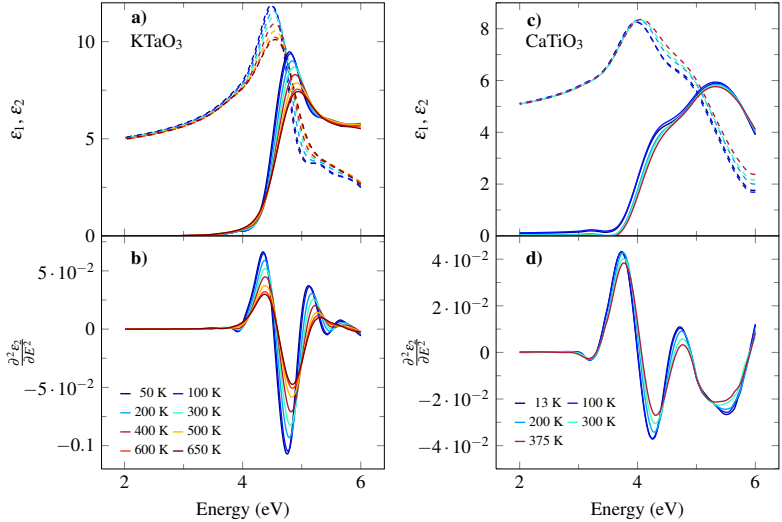

Figure S1: Ellipsometry spectra measured at different temperatures and the corresponding second derivative spectra for KTaO$_3$ (**a)**, **b)**) and CaTiO$_3$ (**c)**, **d)**) samples.

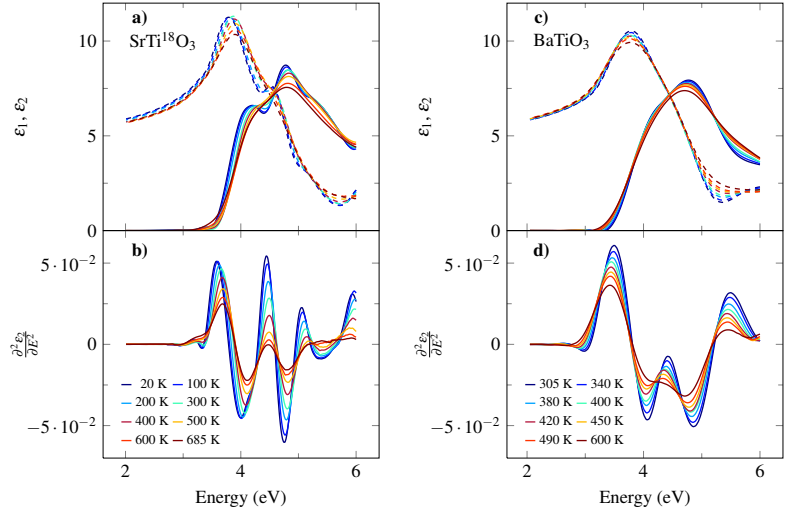

Figure S2: Ellipsometry spectra measured at different temperatures and the corresponding second derivative spectra for SrTi$^{18}$O$_3$ (**a**), **b**)) and BaTiO$_3$ (**c**), **d**)) single crystals.

# Fitting of the direct interband transition in the vicinity of the $M_0$ point in the NIR-UV region

The numerical derivatives of the real and imaginary parts, $\varepsilon_1$ and $\varepsilon_2$, of the measured dielectric function have been independently calculated using an implementation of the Savitzky-Golay smoothing algorithm in Python [2]. The code yields also the $n$-th numerical derivative of the input data with respect to the input $x$-value, i.e. the energy in our case. Care has been taken that no artificial features are introduced by the smoothing procedure. The real and imaginary parts, $\frac{\partial^2 \varepsilon_1}{\partial E^2}$ and $\frac{\partial^2 \varepsilon_2}{\partial E^2}$, have been fitted simultaneously in the vicinity of the $M_0$ critical point using the second derivative of equation (1) given in the paper. After the fitting, the parameters for the energy, $E$, the prefactor, $S$, and the broadening $\Gamma$ for the transition have been obtained as exemplary shown in the main part of the publication for SrTiO$_3$.

It is important to note that the maxima of the derivative spectra do not necessarily coincide with the extracted energies of the interband transitions. The introduction of a phase factor $p = \frac{3\pi}{2}\mathrm{i}$ in the exponent results in a mixing of real and imaginary parts of the dielectric function close to the resonance. This can lead to a small shift of the transition energy away from the observed maxima in the derivative spectra.

## Determination of $\omega_{LO}$

In an insulator like $SrTiO_3$ the eigenfrequency of a longitudinal optical phonon mode, $\omega_{LO}$, can be accurately determined from the zero-crossing of the real part of the dielectric function, $\varepsilon_1$, or likewise from a maximum in the so-called loss function,

$$-\text{Im}\left(\frac{1}{\varepsilon}\right) = \frac{\varepsilon_2}{\varepsilon_1^2 + \varepsilon_2^2} \quad . \tag{1}$$

Figure S3 shows for the case of the room temperature spectra of $SrTiO_3$, $BaTiO_3$, $CaTiO_3$, and $SrTi^{18}O_3$ that this procedure can be used to accurately determine the value of $\omega_{LO}$ (as indicated by the arrows). For these perovskites, a large TO–LO splitting is observed due to the high ionicity of these materials, hence, for the soft mode $\omega_{LO}$ is located in the mid-infrared range at 716, 808, 788, and 765 $cm^{-1}$ for $BaTiO_3$, $CaTiO_3$, $SrTiO_3$, and $SrTi^{18}O_3$, respectively. For $KTaO_3$ we used the value of $\omega_{LO} = 833$ $cm^{-1}$ that is reported in Ref. [3].

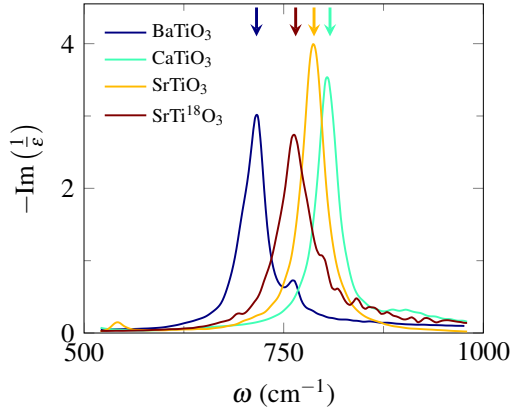

Figure S3: Room temperature loss function of $BaTiO_3$, $CaTiO_3$, $SrTiO_3$, and $SrTi^{18}O_3$ as obtained from the dielectric functions that were measured with mid-infrared ellipsometry. The arrows indicate the LO eigenfrequencies given in the text.

In these insulating titantes, the value of $\omega_{LO}$ of the soft mode is hardly temperature dependent. This is demonstrated in Fig. S4 for the case of $SrTiO_3$ which shows that the position of the peak in the loss function, as derived from our mid-infrared ellipsometry measurements, hardly changes with temperature. This has been also previously shown by Servoin and Gervais in Ref. [4].

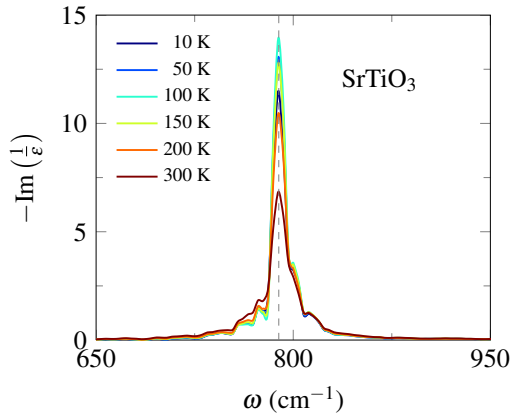

Figure S4: Temperature dependence of the loss function of $SrTiO_3$ as determined from the ellipsometry measurements at different temperatures. The dashed line marks the position of the LO mode energy of $\omega_{LO} = 788$ cm$^{-1}$.

## Soft mode behaviour in perovskite titanates

The eigenfrequency of the transverse optical branch of the soft mode, $\omega_{TO}$, of $SrTiO_3$ is known to decrease from $\sim 95$ cm$^{-1}$ at 300 K to $\sim 15$ cm$^{-1}$ at low temperature [5, 6, 7] due to its quantum paraelectric behavior. Figure S5(a) shows the result of an Lorentz oscillator fit of the soft mode to our far-infrared ellipsometry data over a broad range of temperatures. A similar softening of $\omega_{TO}$ is observed in the incipient-ferroelectric perovskite material $KTaO_3$ and also in $CaTiO_3$ as shown in figure S5(b).

In the ferroelectric state the soft mode is expected to harden again [8]. In the following we show that this behaviour is indeed observed in the ferroelectric state of $SrTi^{18}O_3$ at $T \leq T_{Curie} = 26$ K. Figure S5(c) displays the $T$-dependence of $\omega_{TO}$ which has been deduced from the peak position of the soft mode in the dielectric function. The latter has been obtained from far-infrared reflectivity data of $SrTi^{18}O_3$ by using a Kramers-Kronig transformation. It shows that the soft-mode eigenfrequency, $\omega_{TO}$, exhibits a weak, yet clear minimum at $T_{Curie} = 26$ K as marked by the arrow in figure S5(c). In the ferroelectric phase between 30 and 5 K there is indeed a characteristic hardening of $\omega_{TO}$ by $\sim 5$ cm$^{-1}$. A similar value has been previously obtained from Raman measurements on such $SrTi^{18}O_3$ crystals [9].

The $T$-dependence of the soft mode in $BaTiO_3$ has been extracted from the literature data for the hyper-Raman scattering [10], the far-infrared spectroscopy in Ref. [11] for $T < T_{Curie}$, and the lower energy phonon branch as observed in the infrared spectroscopy

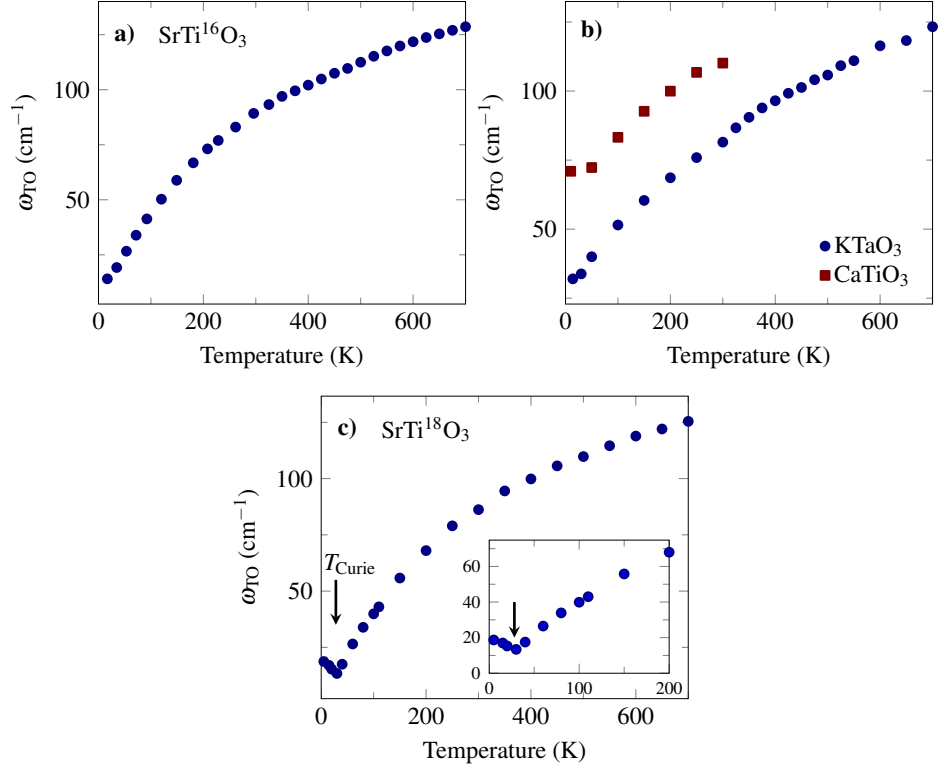

Figure S5: Soft mode eigenfrequency $\omega_{\mathrm{TO}}$ as a function of temperature as obtained from the far-infrared ellipsometry data of the quantum paraelectric materials **a)** SrTiO$_3$ and **b)** KTaO$_3$ and polycrystalline CaTiO$_3$. **c)** Corresponding data for SrTi$^{18}$O$_3$ where the ferroelectric transition at $T_{\mathrm{Curie}} = 26$ K is accompanied by a hardening of the soft mode.

for $T > T_{\mathrm{Curie}}$ in [12]. The estimated phase transition temperature of $T_{\mathrm{Curie}} = 405$ K of these assembled experimental results agrees well with $T_{\mathrm{Curie}}$ of the sample that has been investigated in this work.

## Determination of $\varepsilon_0$ and $\varepsilon_\infty$ from the ellipsometry spectra

For the calculation of the Fröhlich-type interaction according to equation (1) in the main part of the paper, we have used as input the values of $\varepsilon_0$ and $\varepsilon_\infty$. We have derived $\varepsilon_0$ from the Lyddane-Sachs-Teller relation, $\frac{\varepsilon_0}{\varepsilon_\infty} = C \cdot \frac{\omega_{\mathrm{LO}}^2}{\omega_{\mathrm{TO}}^2}$, using the values of $\omega_{\mathrm{TO}}$, $\omega_{\mathrm{LO}}$, and $\varepsilon_\infty$ that have been determined from the far- to mid-infrared ellipsometry spectra. We have

already shown how $\omega_{TO}$ and $\omega_{LO}$ have been obtained from these optical spectra. In the following we show how the value of $\varepsilon_\infty$ has been reliably obtained from these ellipsometry data. Figure S6 shows that at energies well above the phonon range and well below the interband transition region (or the gap energy), the real part of the dielectric function is only very weakly temperature dependent and exhibits a very moderate and continuous dispersion. For our calculations we used the value of $\varepsilon_\infty$ at $E = 0.9$ eV at 300 K. We note that the very weak $T$-variation of $\varepsilon_\infty$ of less than a percent does not have any sizeable effect on our estimate of the band-gap shift due to the Fröhlich interaction as presented in our paper. According to equation (1), the Fröhlich interaction is governed by the very large $T$-dependent changes of $\varepsilon_0$ which arise from the soft-mode behavior.

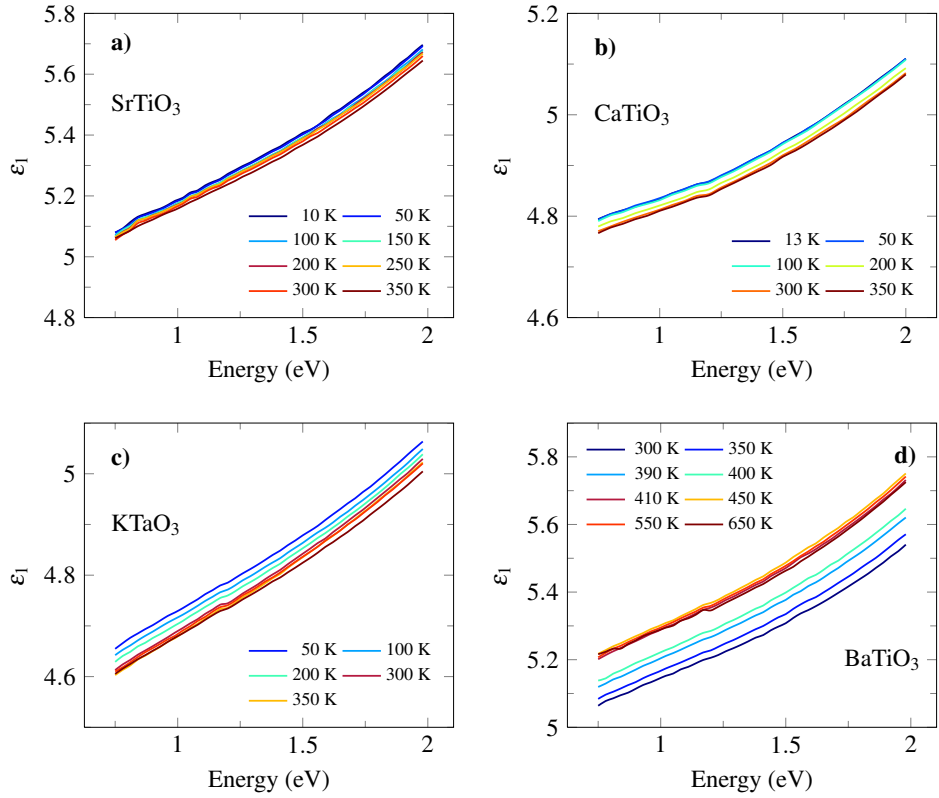

Figure S6: Temperature dependence of the real part of the dielectric function, $\varepsilon_1$, as measured by ellipsometry for **a)** SrTiO$_3$, **b)**, CaTiO$_3$, **c)** KTaO$_3$, and **d)** BaTiO$_3$.

# References

[1] JA Woollam Co., Inc. `http://www.jawoollam.com/`.

[2] SciPy. `http://www.scipy.org/`.

[3] P. A. Fleury and J. M. Worlock. Electric-Field-Induced Raman Scattering in $SrTiO_3$ and $KTaO_3$. *Phys. Rev.*, 174:613–623, 1968.

[4] J. L. Servoin, Y. Luspin, and F. Gervais. Infrared dispersion in $SrTiO_3$ at high temperature. *Phys. Rev. B*, 22:5501–6, 1980.

[5] W. Cochran. Crystal stability and the theory of ferroelectricity. *Adv. Phys.*, 9(36):387–423, 1960.

[6] A. S. Barker and M. Tinkham. Far-Infrared Ferroelectric Vibration Mode in $SrTiO_3$. *Phys. Rev.*, 125:1527–1530, 1962.

[7] H. Vogt. Refined treatment of the model of linearly coupled anharmonic oscillators and its application to the temperature dependence of the zone-center soft-mode frequencies of $KTaO_3$ and $SrTiO_3$. *Phys. Rev. B*, 51(13):8046–8059, Apr 1995.

[8] A Yamanaka, M Kataoka, Y Inaba, K Inoue, B Hehlen, and E Courtens. Evidence for competing orderings in strontium titanate from hyper-Raman scattering spectroscopy. *Europhys. Lett.*, 50(5):688–694, 2000.

[9] M. Takesada, M. Itoh, and T. Yagi. Perfect Softening of the Ferroelectric Mode in the Isotope-Exchanged Strontium Titanate of $SrTi^{18}O_3$ Studied by Light Scattering. *Phys. Rev. Lett.*, 96(22):227602, 2006.

[10] H. Vogt, J. A. Sanjurjo, and G. Rossbroich. Soft-mode spectroscopy in cubic $BaTiO_3$ by hyper-Raman scattering. *Phys. Rev. B*, 26(10):5904–5910, Nov 1982.

[11] J. Hlinka, T. Ostapchuk, D. Nuzhnyy, J. Petzelt, P. Kuzel, C. Kadlec, P. Vanek, I. Ponomareva, and L. Bellaiche. Coexistence of the Phonon and Relaxation Soft Modes in the Terahertz Dielectric Response of Tetragonal $BaTiO_3$. *Phys. Rev. Lett.*, 101(16):167402, Oct 2008.

[12] I. Ponomareva, L. Bellaiche, T. Ostapchuk, J. Hlinka, and J. Petzelt. Terahertz dielectric response of cubic $BaTiO_3$. *Phys. Rev. B*, 77:012102, Jan 2008.


\end{widetext}

\end{document}